\documentclass[sn-mathphys]{sn-jnl}

\jyear{2022}%

\theoremstyle{thmstyleone}%
\theoremstyle{thmstyletwo}%

\theoremstyle{thmstylethree}%

\usepackage{dcolumn}
\usepackage{bm}

\usepackage{color}

\usepackage{mathtools}
\DeclarePairedDelimiter{\abs}{\lvert}{\rvert}

\begin{document}

\title[Stability of Hilda asteroids in three-body problem]{Stability of Hilda asteroids at 3:2 resonance point in restricted three-body problem}


\author*[1]{\fnm{Kosuke} \sur{Asano}}\email{szc04002@st.osakafu-u.ac.jp}

\author[2]{\fnm{Kenichi} \sur{Noba}}\email{noba@omu.ac.jp}
\equalcont{These authors contributed equally to this work.}

\author[3,4]{\fnm{Tomio} \sur{Petrosky}}\email{petrosky@physics.utexas.edu}
\equalcont{These authors contributed equally to this work.}

\affil*[1]{\orgdiv{Department of Physical Science}, \orgname{Osaka Prefecture University}, \orgaddress{\street{Naka-ku}, \city{Sakai}, \postcode{599-8531}, \state{Osaka}, \country{Japan}}}

\affil[2]{\orgdiv{Department of Physics}, \orgname{Osaka Metropolitan University}, \orgaddress{\street{Naka-ku}, \city{Sakai}, \postcode{599-8531}, \state{Osaka}, \country{Japan}}}

\affil[3]{\orgdiv{Center for Complex Quantum Systems}, \orgname{University of Texas}, \orgaddress{\city{Austin}, \postcode{6TX 78712}, \state{Texas}, \country{USA}}}

\affil[4]{\orgdiv{Institute of Industrial Science}, \orgname{University of Tokyo}, \orgaddress{\city{Kashiwa}, \postcode{277-8574}, \state{Chiba}, \country{Japan}}}


\abstract{Stability of Hilda Asteroids in the solar system around the 3:2 resonance point is analyzed in terms of the Sun-Jupiter-asteroid {\it elliptic}  restricted three-body problem. We show that the Hamiltonian of the system is well-approximated by a single-resonance Hamiltonian around the 3:2 resonance. This implies that orbits of the Hilda asteroids are approximately integrable, thus their motion is stable. This is in contrast to other resonances such as the 3:1 and the 2:1 resonances at which Kirkwood gaps occur. Indeed, around the 3:1 and the 2:1 resonances, the Hamiltonians are approximated by double-resonance Hamiltonians that are nonintegrable and thus indicate chaotic motions. By a suitable canonical transformation, we reduce the number of degrees of freedom for the system and derive a Hamiltonian which has two degrees of freedom. As a result, we can analyze the stability of the motion by constructing Poincar\'e surface of section.}

\keywords{Kirkwood gap, Hilda asteroids, elliptic three-body problem, single vs. double resonance Hamiltonian, stability}



\maketitle

\section{Introduction}
\label{backgraund_abstract}

\begin{figure}[b]
  \begin{center}
      \includegraphics[width=7cm,height=5cm,keepaspectratio]{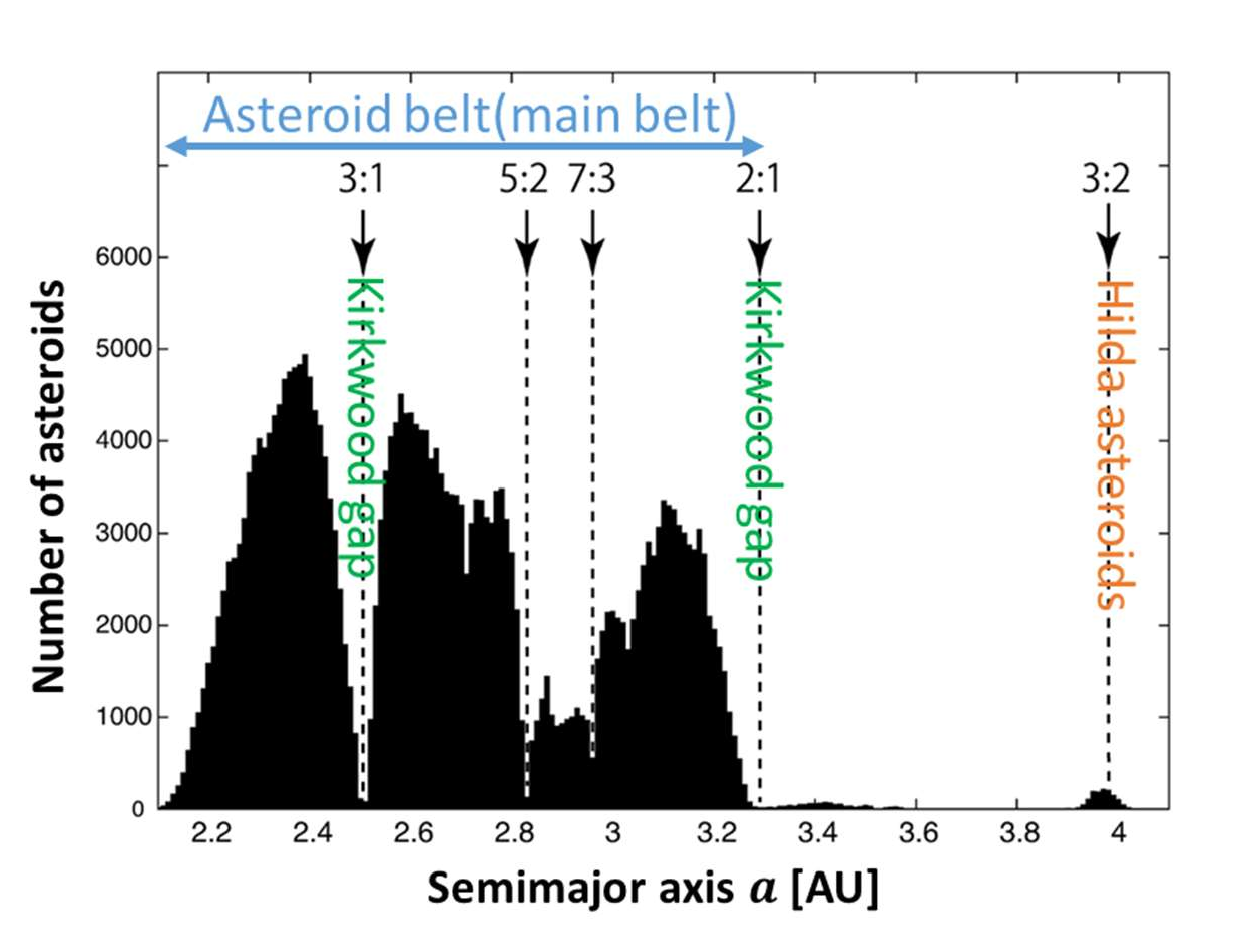}
      \caption{Distribution of the asteroids as a function of the semimajor axis\cite{a}. Kirkwood gaps are observed, for example, at the 3:1 resonance point and the 2:1 resonance point. In contrast,  there are many asteroids at the 3:2 resonance point, which are called as Hilda asteroids}
      \label{fig_Kirkwoodgaps}
  \end{center}
\end{figure}
 In spite of the fact that motions of asteroids inside Kirkwood gaps around the 3:1 resonance and the 2:1 resonance with respect to the orbital period of Jupiter are unstable due to the resonance instability\cite{m}, it has been observed that motions of Hilda asteroids around the 3:2 resonance are stable and form well-known Hilda triangle. The resonance leads to unstable motion in one case, while it leads to stable motion in another case(see Fig.~\ref{fig_Kirkwoodgaps}). How can we understand this apparent contradiction? This problem has been discussed for a long time\cite{ab,n,ac,s,j,ag}, but it seems for us that the reason of the stability of Hilda asteroids is not yet well understood\cite{j,r,ah}. 
 
  In this paper we will show that under a reasonable approximation scheme, one can reduce the Hamiltonian of the  the Sun-Jupiter-asteroid {\it elliptic}  restricted three-body problem to a single-resonance Hamiltonian for the 3:2 resonance case\cite{u,w}. This implies that the Hilda asteroids are approximately integrable, thus, stable\cite{aa,ae}. This is in contrast to other resonances such as the 3:1 and the 2:1 resonances at which Kirkwood gaps occur. Indeed, due to our approximation scheme, the Hamiltonians at the 3:1 and the 2:1 resonances are approximated by double-resonance Hamiltonians that are nonintegrable, and thus indicate chaotic motions\cite{t,ai}.
 
Our approximation scheme consists of two processes. First we will identify small parameters in our interesting domain of the motion. Then, we will use series expansion in terms of these small parameters. We will approximate the Hamiltonian by dominant terms of this series expansion. Then, we will show that the approximated Hamiltonian consists of slow oscillating terms and rapid oscillating terms in time. The Hamiltonian is still complicated to construct analytic solution of the equation of motion, because the Hamiltonian is a time dependent nonlinear Hamiltonian with two degrees of freedom.

 As the next step, we will drop the rapid oscillating terms by taking the time average with an appropriate time scale\cite{ad}.  After this second level of the approximation, we will show that one degree of freedom is reduced in our Hamiltonian with a suitable canonical transformation. This reduction makes it possible for us to  construct two-dimensional Poincar\'e surface of section to analyze stability of motions.  Moreover, we will show that this procedure leads to a single-resonance Hamiltonian for the 3:2 resonance, while it leads to double-resonance Hamiltonians for the 3:1 and the 2:1 resonances. Because a system described by a single-resonance Hamiltonian is integrable, the motion of the asteroids around the 3:2 resonance point is regular and we can find the analytic solution of the equation of motion. On the other hand, a system with the double-resonance Hamiltonian is a nonintegrable system that indicates unstable chaotic motion. Indeed, we will show well-defined regular lines in the Poincar\'e surfaces of section for the case of the 3:2 resonance, while scattered chaotic points in the Poincar\'e surfaces of section for the 3:1 and the 2:1 resonances\cite{l,af}.

We will also numerically compare our solutions of the equation of motion obtained by our approximated Hamiltonian to the ones obtained by the original equation of motion without approximation. Due to a drastic approximation where we drop rapid oscillating contributions even though the magnitude of these terms is comparable to the slow oscillating contributions that are retained in our approximated Hamiltonian, we see some deviation in the time evolution of the eccentricity of the asteroids. Nevertheless, we will see that our analytic solution indicates reasonable shape of Hilda triangle, and also several physical quantities, such as the semimajor axis of the asteroids, show a good agreement.  
				
The structure of this paper is as follows: In Section \ref{three-bodies}, we introduce a Hamiltonian in the elliptic restricted three-body problem. In Section \ref{resonance}, we approximate the Hamiltonian in terms of the small parameters, as well as time averaging procedure for the 3:2, the 3:1 and the 2:1 resonances. In Section \ref{reduction}, we reduce the degree of freedom of the approximate Hamiltonian by a canonical transformation. We show that the Hamiltonian for the 3:2 resonance may be approximately a single-resonance Hamiltonian, while the Hamiltonians for the 3:1 and the 2:1 resonances may be approximately double-resonance Hamiltonians. In Section \ref{poincare32}, we analyze integrability of the motion of the asteroid around the 3:2, the 3:1 and the 2:1 resonances by constructing Poincar\'e surfaces of section. In Section \ref{discussion}, we discuss the validity of our approximation of the Hamiltonian. We summaries our results at Section \ref{concluding}.  We present several appendices to give detail calculations.

\section{Elliptic Restricted Three-body Problem for Asteroid}
\label{three-bodies}

\begin{figure}[b]
\begin{center}
\includegraphics[width=10cm,height=6cm,keepaspectratio]{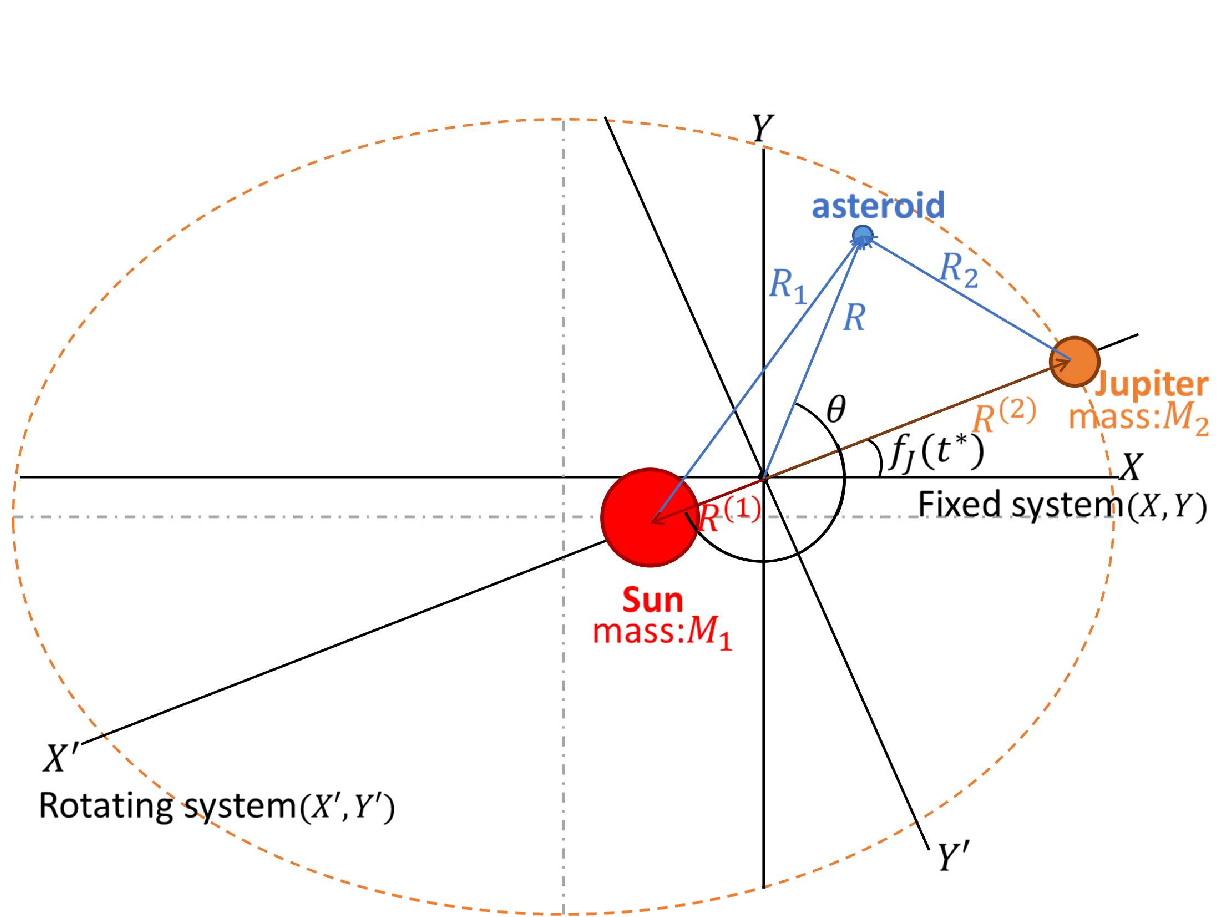}
\caption{The coordinate $(X, Y)$ is the fixed(sidereal) coordinate system. The $X$ axis is parallel to the long axis of the elliptical orbit of Jupiter and $Y$ axis is orthogonal to $X$ axis. The coordinate $ (X', Y') $ is the rotating(synodic) system that rotate with the Sun and Jupiter and $X'$ axis 
}
\label{fig_ellpic}
\end{center}
\end{figure}

In this paper, we analyze the motion of the asteroid as an elliptic restricted three-body problem of the Sun-Jupiter-asteroid in which the Sun and Jupiter (primaries) revolve around their center of mass in elliptic orbits\cite{c}. 
The asteroid is attracted by the primaries. We assume that the asteroid moves in the plane defined by the primaries. 
Since most of the inclinations of real asteroids are distributed less than $10^\circ$ orbital inclinations to the orbit plane of the primaries, it might be a reasonable approximation to get essential feature of the motion of the asteroids (see Fig.~\ref{fig_ellpic}).

The Hamiltonian of the asteroid in the fixed orthogonal coordinate system $(X, Y)$ with the origin at the center of mass between the Sun and Jupiter is given by
\begin{eqnarray}
H'(X,Y,P_X,P_Y,t^*)=\frac{1}{2} (P_X^2+P_Y^2)-G\Big\{\frac{M_1}{R_1(X,Y,t^*)}+\frac{M_2}{R_2(X,Y,t^*)}\Big\}, \hspace{5pt}
\label{eq_ell-three-bodies-hamiltonian_t*_start}
\end{eqnarray}
where $ M_1 $ is the mass of the Sun, $ M_2 $ is the mass of Jupiter, and $ G $ is the gravitational constant. The quantity
$ R_1 (X, Y, t ^ *) $ is the distance from the Sun to the asteroid, and $ R_2 (X, Y, t ^ *) $ is the distance from Jupiter to the asteroid. We use the notation $t^*$ for the time where the origin of a time $t^*=0$ is the time at which the Jupiter passes its perihelion. The momenta $(P_X, P_Y)$ are conjugate to $ (X, Y) $. Note that we put the mass of the asteroid to unity in the Hamiltonian~(\ref{eq_ell-three-bodies-hamiltonian_t*_start}). This is possible without loss of generality because the mass of the asteroid does not influence on the motion of the Sun and Jupiter in the restricted three-body problem\cite{c}.

We use $R$ for the distance between the asteroid and the origin of the coordinate as shown in Fig.~\ref{fig_ellpic}.
We also use the notation  $ f_J (t^*) $ for the true anomaly of Jupiter. In the rotating coordinate system $(X', Y')$ rotating with the primaries, the angle between the $ X $ axis and the $ X'$ axis is given by $ f_J (t^*)- \pi$.

In this paper, we use the following dimensionless units for the time, distance, and mass: For the time, we use a unit in which Jupiter's orbital period $ T_J $ (about 11.86 years) is $ 2 \pi $, for distance, Jupiter's semimajor axis relative to the Sun in the Sun-Jupiter two body problem $ a_J $ (about 5.2 AU) is 1,  and for the mass, the total mass of the Sun and Jupiter, $ M = M_1 + M_2$, is 1.
Due to Kepler's third law,
\begin{equation}
T_J=2 \pi {a_J}^{3/2} (GM)^{-1/2},
\label{eq_tj-aj_dimension}
\end{equation}
the gravitational constant $ G $ is 1 in these dimensionless units. Moreover, $ \mu\equiv M_2/M$ is the mass of Jupiter, and $ 1 -\mu =M_1/M$ is the mass of the Sun. In the dimensionless units, we use the notations $ t $, $ x $, $ y $, $p_x$, $p_y$, $ r_1 $, $ r_2 $, $r$ for $t^*, X, Y, P_X, P_Y, R_1, R_2, R$, respectively.

Performing  the canonical transformation from the fixed orthogonal coordinate system $ (x, y) $ to the rotating polar coordinate system $ (r, \theta) $,
\footnote{
In order to derive the Hamiltonian~(\ref{eq_ell-three-bodies-hamiltonian_t}) in the rotating polar coordinate system, we perform the following two-step canonical transformations with the generating functions by introducing intermediate variables with the tilde notations:
\begin{eqnarray}
W_1(p_{x},p_{y},\tilde{r},\tilde{\theta})
   &=-\tilde{r}(p_{x}\cos {\tilde{\theta}}+p_{y}\sin {\tilde{\theta}}), \nonumber \\
W_2(\tilde{r} ,\tilde{\theta} ,p_r,p_{\theta} ,t)
   &=\tilde{r}p_r+\big(\tilde{\theta}-f_J(t)+\pi \big)p_{\theta}. \nonumber
\end{eqnarray}
}
we obtain the Hamiltonian,
\begin{eqnarray}
H_t(r,\theta,p_r,p_{\theta},t)&=&\frac{1}{2}\Big (p_{r}^2+\frac{p_{\theta}^2}{r^2}\Big)-F_t(f_J(t))^{-1}p_{\theta} \nonumber \\
& &-\frac{1-\mu}{r_1(r,\theta,f_J(t))}-\frac{\mu}{r_2(r,\theta,f_J(t))}, \hspace{5pt}
\label{eq_ell-three-bodies-hamiltonian_t}
\end{eqnarray}
\begin{eqnarray}
r_1(r,\theta,f_J(t))&=&\sqrt{r^2+\big\{r^{(1)}(f_J(t))\big\}^2-2rr^{(1)}(f_J(t))\cos {\theta}}, \\
r_2(r,\theta,f_J(t))&=&\sqrt{r^2+\big\{r^{(2)}(f_J(t))\big\}^2+2rr^{(2)}(f_J(t))\cos {\theta}},
\end{eqnarray}
\begin{eqnarray}
r^{(1)}(f_J(t))&=&\mu r^0(f_J(t)), \\
r^{(2)}(f_J(t))&=&(1-\mu) r^0(f_J(t)),
\end{eqnarray}
where $p_r$ and $p_\theta$ are the momenta conjugate to $r$ and $\theta$, respectively, $r^{(1)}$ is the distance from the origin to the Sun, and $r^{(2)}$ is the distance from the origin to Jupiter. Moreover, $r^0$ is the distance from the Sun to Jupiter in the Sun-Jupiter two-body problem,
\begin{equation}
r^0(f_J(t)) =\frac{1-e_J^2}{1+e_J\cos{f_J(t)}}, \label{eq_r^0}
\end{equation}
with  the eccentricity of Jupiter $e_J$, and $F_t(f_J(t))$ is defined by
\begin{equation}
F_t(f_J(t))\equiv  \frac{1}{(d f_J/d t)}  =\frac{\big\{r^0(f_J(t))\big\}^2}{\sqrt{1-e_J^2}}. \label{eq_Ft}
\end{equation}

The true anomaly $ f_J $  of Jupiter is a complicate function of time $t$, since for example we have the relation,
\begin{eqnarray}
\cos f_J(t)&=&-e_J+\frac{2(1-e_J^2)}{e_J}\sum_{n'=1}^{\infty}J_{n'}(n'e_J)\cos {(n't)} \nonumber \\
&=&\cos t+e_J(\cos{2t}-1)+\frac{9}{8}e_J^2(\cos{3t}-\cos t)+O(e_J^3), \label{eq_cosfj-ejl}
\end{eqnarray}
where $J_{n'}(z)$ is a Bessel function of the first kind \cite{f}. 

In order to avoid unnecessary complication in the following calculation, we will use $ f_J $ as a {\it new time}.  This is possible, because the time $ t $ always appears through the function $ f_J (t) $ in the Hamiltonian\cite{c,v}.

By introducing the new Hamiltonian,
\begin{eqnarray}
H_{f_J}(r,\theta,p_r,p_{\theta},f_J) \hspace{-60pt} \nonumber \\
&\equiv& F_t(f_J)H_t(r,\theta,p_r,p_{\theta},t(f_J))
\label{eq_hamiltonian_tfjtranspose} \\
&=&\frac{1}{2}\Big (p_{r}^2+\frac{p_{\theta}^2}{r^2}\Big)F_t(f_J)-p_{\theta} \nonumber \\
& &+\Big\{ -\frac{1-\mu}{r_1(r,\theta,f_J)}-\frac{\mu}{r_2(r,\theta,f_J)}\Big\} F_t(f_J),
\label{eq_ell-three-bodies-hamiltonian_fj}
\end{eqnarray}
we obtain a new set of the Hamilton equations of motion,
\begin{eqnarray}
\frac{dr}{df_J}
    &=& \frac{\partial H_{f_J}}{\partial p_r}, \qquad \frac{dp_r}{df_J}=-\frac{\partial H_{f_J}}{\partial r}, \nonumber \\
 \frac{d \theta}{df_J} 
    &=& \frac{\partial H_{f_J}}{\partial p_{\theta}},\qquad \frac{d p_{\theta}}{df_J} = -\frac{\partial H_{f_J}}{\partial \theta}.
\label{eq_new_HamiltonEq}
\end{eqnarray}
Hereafter we refer to $ f_J $ as time.

We divide the Hamiltonian into an unperturbed term and perturbation terms as
\begin{eqnarray}
H_{f_J}(r,\theta,p_r,p_{\theta},f_J)
   =H'_0 + V_{e_J} + \mu V_{\rm e}' , \label{eq_ell-three-bodies-hamiltonian_fj_sepa}
\end{eqnarray}\vspace{-20pt}
\begin{eqnarray}
H'_0
   &=&\frac{1}{2}\Big (p_{r}^2+\frac{p_{\theta}^2}{r^2}\Big)-p_{\theta}-\frac{1}{r}, \label{eq_ell-three-bodies-hamiltonian_fj_sepa_non-perturbation} \\
V_{e_J}
   &=&\Big\{ \frac{1}{2}\Big (p_{r}^2+\frac{p_{\theta}^2}{r^2}\Big)-\frac{1}{r}\Big\} \Big( F_t(f_J)-1 \Big), \label{eq_ell-three-bodies-hamiltonian_fj_sepa_vej} \\
\mu V_{\rm e}'
   &=&\Big\{ \frac{1}{r}-\frac{1-\mu}{r_1}-\frac{\mu}{r_2}\Big\} F_t(f_J).
\label{eq_ell-three-bodies-hamiltonian_fj_sepa_muvep}
\end{eqnarray}
Here $ H_0'$ is the unperturbed  Hamiltonian  which is the Hamiltonian in the Sun-asteroid two-body problem for $ \mu = 0 $ and $ e_J = 0 $. The interaction part $ V_{e_J} $ is the perturbation due to  the non-uniform rotation of the rotating coordinate system to the fixed coordinate system.  The perturbation $ V_{e_J} $ is independent of $ \mu $, while it depends on $e_J$ and vanishes for $e_J=0$.
The interaction part $ \mu V_{\rm e}' $ is the perturbation coming from the gravitational force of Jupiter to the asteroid.
The Hamiltonian~(\ref{eq_ell-three-bodies-hamiltonian_fj_sepa}) is a periodic time $ f_J $ dependent Hamiltonian with two degrees of freedom .
 
 We further perform a canonical transformation of the Hamiltonian in terms of the Delaunay variables $(l, g, L, G)$. Then we obtain
\begin{eqnarray}
H_{f_J}'(L,G,l,g,f_J)= H'_0 + V_{e_J}+ \mu V_{\rm e}', \label{eq_ell-three-bodies-hamiltonian_fj_sepa_lg}
\end{eqnarray}\vspace{-20pt}
\begin{eqnarray}
H'_0
    &=&-\frac{1}{2L^2}-G, \label{eq_ell-three-bodies-hamiltonian_fj_sepa_lg_non-perturbation} \\
V_{e_J}
   &=&-\frac{1}{2L^2}\Big( F_t(f_J)-1 \Big), \label{eq_ell-three-bodies-hamiltonian_fj_sepa_lg_vej} \\
\mu V_{\rm e}' &=& \mu \hspace{-7pt} \sum_{m',n',\alpha'} \hspace{-7pt} V_{{\rm e} ,m',n',\alpha'}(L,G;e_J,\mu)e^{i(m'g+n'l+\alpha' f_J)},\hspace{10pt} \label{eq_ell-three-bodies-hamiltonian_fj_sepa_lg_muvep2}
\end{eqnarray}
where the summation on $m',n'$ and $\alpha'$ in the  Fourier series of $V_{{\rm e}}' $ are taken over all integers. (Note this is  a $ 2 \pi $ periodic function with respect to the angles $ g $, $ l $ and $ f_J$.)

 The relations between  
  the variables  $ (r, \theta, p_r, p_ {\theta}) $ and the Delaunay variables ($ l, g, L, G $) are given by
\begin{eqnarray}
L&=&\sqrt{a}, \label{eq_L-a} \\
G&=&\sqrt{a(1-e^2)}=p_{\theta}, \label{eq_G-ae} \\
r&=&a(1-e\cos u) ,\label{eq_r-aeu} \\
l&=&u-e\sin u ,\label{eq_keplar-eq} \\
r&=&\frac{a(1-e^2)}{1+e\cos {(\theta-g)}}=\frac{a(1-e^2)}{1+e\cos {f}}, \label{eq_r-aef} \\
g&=&\theta-f ,\label{eq_g-thetaf} \\
p_r&=&\sqrt{-\frac{G^2}{r^2}+\frac{2}{r}-\frac{1}{L^2}} \label{eq_pr-rLG},
\end{eqnarray}
where $ a $ is the semimajor axis,  $ e $ is the eccentricity, $ u $ is the eccentric anomaly, $ f $ is the true anomaly, $ l $ is the mean anomaly, and $ g $ is the longitude of the apsidal line in the rotating coordinate system which are all associated to asteroid\cite{f}.  
We have the relations,
\begin{eqnarray}
a&=&L^2, \label{eq_a-L} \\
e&=&\sqrt{1-(G/L)^2}. \label{eq_e-LG}
\end{eqnarray}

Note that Eq.~(\ref{eq_ell-three-bodies-hamiltonian_fj_sepa_lg_non-perturbation}) is the Hamiltonian for the two-body problem, and $L$, $G$, $a$ and $e$ are constants of motion in the two-body problem. Moreover, for two-body problem, we can obtain
\begin{eqnarray}
l&=& a_{mn}^{-\frac{3}{2}}t+l_0=\frac{m}{n}t+l_0, \label{eq_two-bodies_lt} \\
g&=&-t+g_0 , \label{eq_two-bodies_gt}
\end{eqnarray}
at the $m:n$ resonance point, where $l_0$ and $g_0$ are the initial conditions of $l$ and $g$, respectively, and $a_{mn}=(n/m)^{2/3}$ is the semimajor axis of the asteroid at the $m:n$ resonance point which is obtained by Kepler's third law at the $m:n$ resonance,
\begin{equation}
\frac{(\frac{n}{m}\times T_J)^2}{a_{mn}^3}=\frac{T_J^2}{a_J^3},
\label{keplar-3_sun-asteroid}
\end{equation}
where $a_J$ is equal to 1 in this dimensionless unit. Because of Eqs.~(\ref{eq_two-bodies_lt}) and (\ref{eq_two-bodies_gt}), $mg+nl$, $g+t$ and their linear combination angles are constants at the $m:n$ resonance in the two-body problem.

\section{Approximation of Perturbation Terms around $m:n$ Resonance Point}
\label{resonance}

Since the restricted three-body problem is nonintegrable in the sense of Poincar\'e, it is hopeless to construct  exact solution of the equation of motion. 
Thus, in order to understand the behavior of the asteroid through the equations of motion, we perform two kinds of approximations on the Hamiltonian: 

(1) the usual approximation in which we expand the Hamiltonian~(\ref{eq_ell-three-bodies-hamiltonian_fj_sepa_lg}) in terms of small quantities,

(2)  we neglect highly oscillating terms by taking a long time average.

\noindent
We first note that our dimensionless fixed physical parameters have small values, i.e., $\mu \approx 1.0\times 10^{-3}$, and $e_J \approx 0.05$. Hence, we expand our Hamiltonian in the series of these parameters. We then neglect terms of order $\mu^2$ and terms with combination of $\mu$ and $e_J$ that are the same order as $\mu^2$. As will be explained below, this procedure is performed  depending on the type of resonance called as $(m-n)$th-order resonance. Here we use the terminology $(m-n)$th-order resonance for the $m:n$ resonance to specify the type of resonance. For example,  the 3:2 resonance and the 2:1 resonance correspond to the first-order resonance, while the 3:1 resonance corresponds to the second-order resonance, respectively. Recall that the 3:1 and the 2:1 resonances corresponding to the Kirkwood gaps, while the 3:2 resonance corresponding to the Hilda asteroids. 

Moreover, we note that most of the asteroids of our interest have the valuables of $e$ and $a$ as
$e < 0.4$ and $a < 0.8$. Hence, we also approximate the Hamiltonian up to appropriate order of $e$ and $a$ which is consistent to the approximation with  neglecting terms of order $\mu^2$.

In order to perform the approximation based on these small quantities, it is convenient to represent the perturbation term  $\mu V_{\rm e}'$  by the valuables $a$, $e$,  $\theta$, and $f$, instead of the Delaunay variables.  Following these approximation scheme, we expand the Hamiltonian in terms of  $\mu $ and $e_J $, and also expand $1/r_1$ and $1/r_2$ in the series of $\mu r^0/r$ and $r/(1-\mu) r^0$\cite{d,g}. Then, we obtain for $\mu V_{\rm e}'$,
\begin{eqnarray}
\mu V_{\rm e}'=
  &\mu& {\{a(1-e^2)\}}^{-1}(1+e\cos f) \Big\{ 1-2e_J \cos f_J+\frac{3}{2}e_J^2\cos{2f_J}+O(e_J^3)\Big\} \nonumber \\
  &-&\mu \{a(1-e^2)\}^{-2}(1+e \cos f)^2 \cos{\theta} \nonumber \\
  & & \times \Big[ 1-3e_J \cos f_J+e_J^2\Big\{ \frac{1}{2}+3\cos{2f_J}\Big\}+O(e_J^3) \Big] \nonumber \\
  &-&\mu \Big\{ 1-e_J \cos f_J+\frac{1}{2}e_J^2\cos{2f_J}+O(e_J^3) \Big\}  \nonumber \\
  &+&\mu \sum_{i=1}^{\infty}\sum_{j=0}^{\infty}\sum_{k=0}^{i-1} P_i(\cos{\theta}) (-1)^{i+j+1} \hspace{3pt} _{i+j-1}{\rm C}_{j} \hspace{3pt} _{i-1}{\rm C}_{k} \nonumber \\
& & \times \{a(1-e^2)\}^i (e \cos f)^j (e_J \cos f_J)^k \nonumber \\
&+&O(\mu^2),
\label{eq_muvep_expansion_ftheta}
\end{eqnarray}
where $P_{i}(x)$ in Eq.~(\ref{eq_muvep_expansion_ftheta})  is the $i$th order Legendre polynomial, and $ _{i}{\rm C}_{j}$ is the binomial coefficient. Here we have used Eq.~(\ref{eq_r-aef}) and the expansion forms,
\begin{eqnarray}
(1\pm e \cos f)^{i}&=&\sum_{j=0}^{i} \hspace{3pt} _{i}{\rm C}_{j}(\pm e \cos f)^j \hspace{5pt} (i=1,2,...),\hspace{5pt} \label{eq_app_1-ejcosfj} \\
(1\pm e \cos f)^{-i}&=&\sum_{j=0}^{\infty} \hspace{3pt} _{i+j-1}{\rm C}_{j} (\mp e \cos f)^{j}. \label{eq_app_1+-ecosf}
\end{eqnarray}
We will express this interaction part in terms of the Delaunay variables, later.

Similarly, for $V_{e_J}$, we obtain the perturbation term 
\begin{eqnarray}
V_{e_J} &=& -\frac{1}{2L^2}\Big\{ \sum_{i=0}^{\infty} (1-e_J^2)^{\frac{3}{2}}(i+1)(-e_J \cos{f_J})^i -1 \Big\} \nonumber \\
&=& -\frac{1}{2L^2}\Big\{ -2e_J\cos f_J +\frac{3}{2}e_J^2\cos 2f_J -e_J^3\cos 3f_J  \nonumber \\
& & +e_J^4\Big(\frac{1}{4}\cos 2f_J+\frac{5}{8}\cos4f_J \Big)-\frac{3}{8}e_J^5(\cos 3f_J+\cos 5f_J)+O(e_J^6) \Big\}. \hspace{10pt}
\label{eq_vej_expansion}
\end{eqnarray}
We note that each terms of the expansion of $V_{e_J}$ depend explicitly on the time $f_J$.

Our Hamiltonian is still too complicate to analyze the equations of motion. We now introduce a drastic approximation.  When we rewrite  $\mu V_{\rm e}'$ with the Delaunay variables, we see a several combinations of the angles $g$, $l$ and the time $f_J$. We first note that $mg+nl$ and $g+t$ are constants in time at the $m:n$ resonance point in the two-body problem as mentioned in the previous section. Hence, for the case $\mu \ll 1$, one can expect that a linear combination of $mg+nl$ and $g+f_J$ is important. 

\begin{figure*}[btph]
  \begin{center}
    \begin{tabular}{c}
        \begin{minipage}{0.45\hsize}
        \begin{center}
          \includegraphics[height=4.0cm,keepaspectratio]{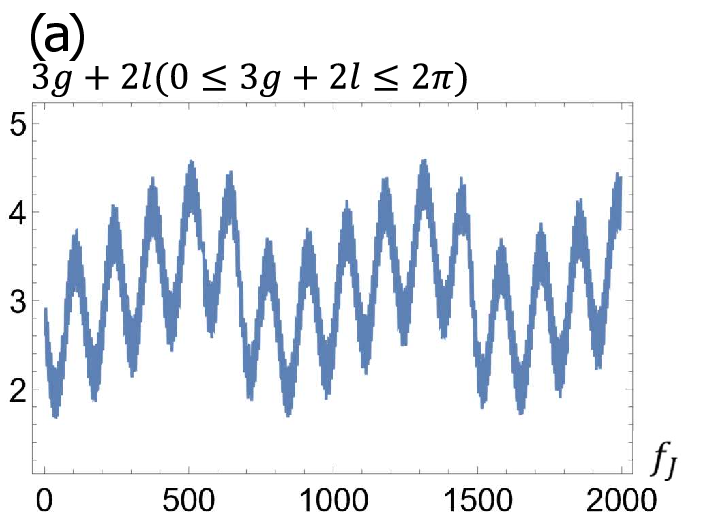}
        \end{center}
      \end{minipage}
        \begin{minipage}{0.45\hsize}
        \begin{center}
          \includegraphics[height=4.0cm,keepaspectratio]{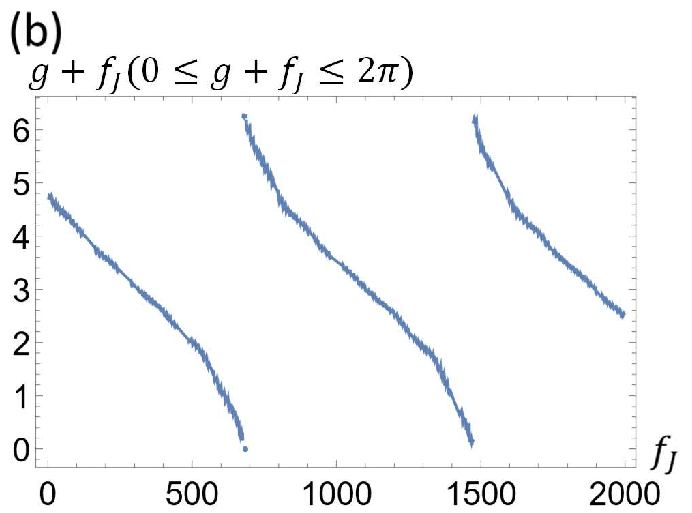}
        \end{center}
      \end{minipage}\\ \\
        \begin{minipage}{0.45\hsize}
        \begin{center}
          \includegraphics[height=3.8cm,keepaspectratio]{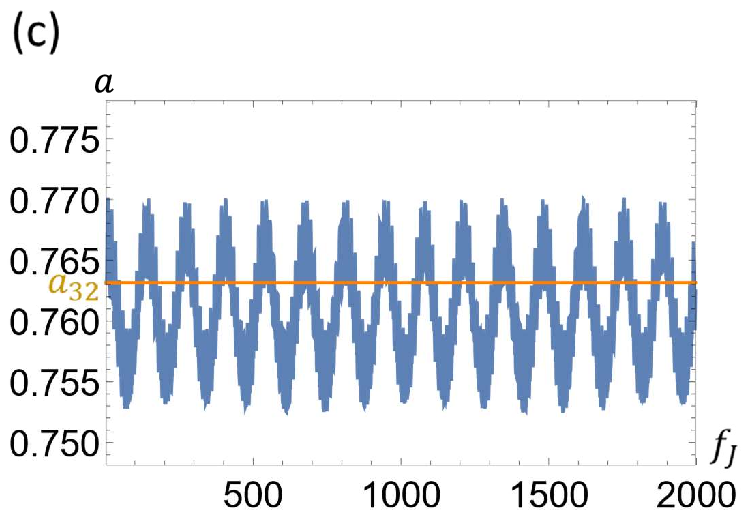}
        \end{center}
      \end{minipage}
        \begin{minipage}{0.45\hsize}
        \begin{center}
          \includegraphics[height=3.8cm,keepaspectratio]{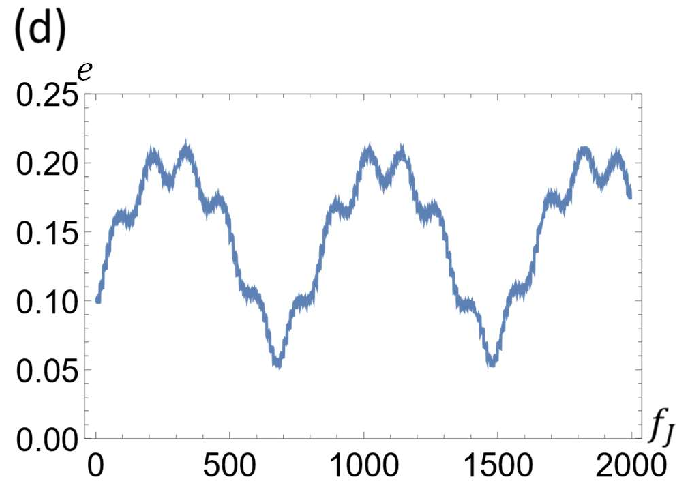}
        \end{center}
      \end{minipage}
    \end{tabular}
\caption{Numerical results of time evolution of the angles $3g+2l$, $g+f_J$, the semimajor axis $a$ and the eccentricity $e$ of the asteroid with the time $f_J$ obtained by the Hamiltonian system~(\ref{eq_ell-three-bodies-hamiltonian_fj}) around the 3:2 resonance point. These results are obtained with initial conditions $(a, e, l, g, f_J)=(a_{32}, 0.10, 2\pi/3, \pi/2, \pi)$. We can see two long periodicities of $3g+2l$ and $g+f_J$, 133 and 798}
\label{fig_ell_time-evolution}
  \end{center}
\end{figure*}

 Indeed, as shown in Fig.~\ref{fig_ell_time-evolution}, a numerical calculation of the original equations of motion for $m=3$ and $n=2$ (the Hilda asteroids) shows that time evolution of $3g+2l$ and $g+f_J$ have long periodicities.
These results have been obtained with the initial conditions $(a, e, l, g, f_J)=(a_{32}, 0.10, 2\pi/3, \pi/2, \pi)$,  where $a_{32}$ is the semimajor axis of the asteroid at the $3:2$ resonance in Eq.~(\ref{keplar-3_sun-asteroid}). As we can see in Figs.~\ref{fig_ell_time-evolution} (a) and (b), the long periods for $3g+2l$ (an oscillation period) and $g+f_J$ (a rotation period) are 133 (about 251 years) and 798 (about 1506 years), respectively. We can also see these periodicities in $a$ and $e$ in Figs.~\ref{fig_ell_time-evolution}(c) and (d).

 We expect that short-period term does not influence the stability of the long-time evolution of the asteroid. Thus, we keep only the terms that contain the linear combination of $mg+nl$ and $g+f_J$ for the $m:n$ resonance case, as well as the angle-independent term $\mu V_{{\rm e} ,0,0,0}$ in Eq.~(\ref{eq_ell-three-bodies-hamiltonian_fj_sepa_lg_muvep2}).
We neglect all other combinations of the angles with short-period terms, in spite of the fact that short-period terms can have the same order of magnitude as the long-period terms with respect to the small quantities $\mu$, $e_J$, $e$ and $a$.

Under this approximation we keep only the terms with the form,
\begin{eqnarray}
  f''_{m'', n''}(L,G,g,l,f_J) \equiv \mu V''_{m'', n''}(L, G) \cos [ m''(mg+nl)+n''(g+f_J) ], 
 \label{eq_angle_dependent_terms} 
\end{eqnarray}
 where $m''=0,1,2,...$ and $n''=...,-1,0,1,2,...$, as well as the $\mu V_{{\rm e} ,0,0,0}$.
This procedure corresponds to taking a time-average of the evolution of motion.  This approximation leads to somewhat rough estimation of the time evolution of the asteroid. However, we expect that this approximation gives a reasonable criteria  of the stability analysis for the long time evolution of the asteroid.

Then, by ignoring fourth-order terms in $e_J$ and $e$ and fourth-order terms in $e_J$ and $a$, we obtain the following approximated Hamiltonian for the 3:1, the 2:1 and the 3:2 resonance cases:
\begin{eqnarray}
\mu V_{\rm e}' \approx& &\hspace{-10pt} \mu a^{-1}(1+e\cos f+e^2+e^3\cos{f}) \nonumber \\
     & &-2\mu e_J a^{-1}(1+e\cos f+e^2)\cos f_J +\frac{3}{2}\mu e_J^2 a^{-1}(1+e\cos{f})\cos{2f_J} \nonumber \\
     &-&\mu a^{-2}\Big\{1+2e \cos f+\frac{e^2}{2}(5+\cos{2f})+4e^3\cos{f}\Big\} \cos{\theta} \nonumber \\
     & &+3\mu e_J a^{-2}\Big\{1+2e \cos f+\frac{e^2}{2}(5+\cos{2f})\Big\} \cos{\theta}\cos f_J \nonumber \\
     & &-\mu e_J^2 a^{-2}(1+2e\cos{f})\cos{\theta} \Big\{ \frac{1}{2}+3\cos{2f_J}\Big\} \nonumber \\
     &-&\mu \Big\{ 1-e_J \cos f_J+\frac{1}{2}e_J^2\cos{2f_J} \Big\}  \nonumber \\
     &+&\mu P_1(\cos{\theta})a^1 \Big\{1-e \cos f - \frac{1}{2}e^2(1-\cos{2f})+\frac{1}{4}e^3(\cos{f}-\cos{3f})\Big\} \nonumber \\
     &+&\mu P_2(\cos{\theta})a^2 \nonumber \\
     & &\times \Big\{-1+2 e \cos f + \frac{1}{2}e^2(1-3\cos{2f})-e^3(\cos{f}-\cos{3f})\Big\}  \nonumber \\
     &+&\mu e_J P_2(\cos{\theta})a^2 \Big\{-1+2 e \cos f + \frac{1}{2}e^2(1-3\cos{2f})\Big\} \cos f_J \nonumber \\
     &+&\mu P_3(\cos{\theta})a^3 \Big\{1-3 e \cos f + 3e^2\cos{2f}+\frac{1}{2}e^3(3\cos{f}-5\cos{3f})\Big\}.  \nonumber \\
\label{eq_muvep_expansion_ftheta_app} 
\end{eqnarray}
This expression still contains angle-independent terms and angle-dependent terms with the type Eq.~(\ref{eq_angle_dependent_terms}), as well as angle-dependent terms without the type Eq.~(\ref{eq_angle_dependent_terms}).

In order to go to the next stage of the calculation we note that the above approximation depends on the $(m-n)$th order resonance. Indeed, in order to be consistent with our approximation where we neglect $\mu^2$ contribution, we should keep the terms up to $(m-n+1)$th order in $e_J$ and $e$ for  $\mu V_{\rm e}'$. Hence, the terms which we should keep in the approximation are different, depending on the first-order resonance (i.e., $m-n=1$), or the second-order resonance (i.e., $m-n=2$). Namely, we have to keep the contributions of  the orders with $\mu e_J^{o_1} e^{(m-n+1)-o_1}$ for $o_1 = 0,1,...,(m-n+1)$  which are greater than  $\mu^2$. Similarly, 
we also have to keep the contributions of  the orders with $\mu e_J^{o_2} a^{m-o_2}$ for $o_2 = 0,1,..,m$.

For example, in the case of the first-order resonance with the 2:1 ($m=2$) or the 3:2 ($m=3$) resonance point, we take into account the perturbation terms which are the same order as or lower order than $\mu e_J^{o_1} e^{2-o_1}$ for $o_1 = 0,1,2$ and $\mu e_J^{o_2} a^{m-o_2}$ for $o_2 = 0,1,...,m$. Thus, only a part of Eq.~(\ref{eq_muvep_expansion_ftheta_app}) contribute for the first-order resonance. On the other hand, all terms in Eq.~(\ref{eq_muvep_expansion_ftheta_app}) contribute for the second-order resonance with the 3:1 resonance.

However, we must be careful when identifying the small-parameter dependence of the expression in Eq.~(\ref{eq_muvep_expansion_ftheta_app}).  Indeed, we are calculating the equations of motion in terms of the Delaunay variables, and not for the variables $a$, $e$, $\theta$ and $f$. For example, in $\cos f$ or  $\cos \theta$ there appear terms involving extra $e$ as
\begin{eqnarray}
\theta
   &=&g+f, \label{eq_theta-g+f}\\
\cos f
   &=&-e+\frac{2(1-e^2)}{e}\sum_{n'=1}^{\infty}J_{n'}(n'e)\cos (n'l) \nonumber \\
   &=&\cos l+e(\cos{2l}-1) +\frac{9}{8}e^2(\cos{3l}-\cos l)+O(e^3), \label{eq_cosf-el}\\
\sin f
   &=&2\sqrt{1-e^2}\sum_{n'=1}^{\infty} \frac{dJ_{n'}(z)}{dz} \mid_{z=n'e} \sin (n'l) \nonumber \\
   &=&\sin l+e\sin{2l} +e^2\Big( \frac{9}{8}\sin{3l}-\frac{7}{8}\sin l\Big) +O(e^3). \label{eq_sinf-el}
\end{eqnarray}
Hence, the above mentioned criteria of the different small-parameter dependence on the $(m-n)$th-resonance should be applied after we express the Hamiltonian in terms of the Delaunay variables. Detailed calculations to identify the small-parameter dependence are presented in  Appendix \ref{section_d}.

In addition, we must also evaluate the contribution from  $V_{e_J}$ in  Eq.~(\ref{eq_vej_expansion}) to the long-time motion. This is simple, because the approximation consistent with Eq.~(\ref{eq_muvep_expansion_ftheta_app}) for $V_{e_J}$ is given by
\begin{eqnarray}
V_{e_J} \approx &-&\frac{1}{2L^2}\Big\{ -2e_J\cos f_J +\frac{3}{2}e_J^2\cos 2f_J -e_J^3\cos 3f_J \nonumber \\
& & +e_J^4\Big(\frac{1}{4}\cos 2f_J+\frac{5}{8}\cos4f_J \Big)\Big\}, 
\label{eq_vej_expansion_app}
\end{eqnarray}
where the fifth order in $e_J$ in Eq.~(\ref{eq_vej_expansion}) has been neglected since $e_J^5  <\mu^2$. Hence, $V_{e_J}$ consists only of $f_J$-depending terms that are highly oscillating terms, and there are no terms with the combination of the angles $mg+nl$ or $g+f_J$ that are slowly oscillating terms. As a result, we can neglect $V_{e_J}$ because it does not influence on the stability of the motion in the long-time motion.

Combining all calculations mentioned above, we finally obtain our approximated form of the angle-dependent part of the interaction $\mu V_{{\rm e},m,n}'$ in Eq.~(\ref{eq_ell-three-bodies-hamiltonian_fj_sepa_lg_muvep2}) (see Appendix \ref{section_d}). This is given by
\begin{eqnarray}
\mu V_{{\rm e},m,n}'(L,G,l,g,f_J;e_J,\mu) \hspace{-100pt} \nonumber \\
    &\approx& \mu C_{mn}'(L,G) \cos [mg+nl] \nonumber \\
    & &+\mu e_J D_{mn}'(L,G) \cos [mg+nl-(g+f_J)] \nonumber \\
    & &+\mu e_J^2 E_{mn}'(L,G) \cos [mg+nl-2(g+f_J)], \hspace{5pt}
\label{myuvemn}
\end{eqnarray} 
with
\begin{eqnarray}
C_{mn}'(L,G)&\equiv& c_{mn}a^m e^{m-n}, \label{cpmn} \\
D_{mn}'(L,G)&\equiv&
  \begin{cases}
    \frac{3}{2}a^{-2} &\hspace{-50pt} \text{for the 2:1 resonance case} \\
    d_{mn}a^{m-1} e^{m-n-1} \\ &\hspace{-50pt} \text{for the other resonance case,}
  \end{cases} \label{dpmn} \\
E_{mn}'(L,G)&\equiv&
  \begin{cases}
   0 &\hspace{-50pt} \text{for the first-order resonance} \\
   -\frac{3}{2}a^{-2} &\hspace{-50pt} \text{for the 3:1 resonance case} \\
   e_{mn}a^{m-2} e^{m-n-2} \\ &\hspace{-50pt} \text{for the other second-order resonance,}
  \end{cases} \label{epmn} 
\end{eqnarray}
where $a$ and $e$ are the functions of $L$ and $G$(see Eqs.~(\ref{eq_a-L}) and (\ref{eq_e-LG})), and
\begin{eqnarray}
c_{mn}
  &\equiv& \sum_{r=0}^{m-n} c_{mn}^{(r)}, \label{cmnsum} \\
d_{mn}
  &\equiv& \sum_{r=0}^{m-n-1} d_{mn}^{(r)}, \label{dmnsum} \\
c_{mn}^{(0)}
  &\equiv& (-1)^{n+1}\hspace{3pt} _{2m-n-1}{\rm C}_{m-1} \frac{p_{\theta,m}}{2^{m-n}}, \label{cmn0} \\
c_{mn}^{(1)}
  &\equiv&
  \begin{cases}
    (-1)^{n+1}\hspace{3pt} _{2m-n-2}{\rm C}_{m-1} \frac{m p_{\theta,m}}{2^{m-n-1}} \\ &\hspace{-100pt} \text{for the first-order resonance} \\
    (-1)^{n+1}\hspace{3pt} _{2m-n-2}{\rm C}_{m-1} \frac{(n+1) p_{\theta,m}}{2^{m-n-1}} \\ &\hspace{-100pt} \text{for the second-order resonance,}
  \end{cases} \label{cmn1} \\
c_{mn}^{(2)}
  &\equiv& (-1)^{n+1} p_{\theta,m} C_{f,m}, \label{cmn2} \\
d_{mn}^{(0)}
  &\equiv& (-1)^{n+1}\hspace{3pt} _{2m-n-3}{\rm C}_{m-2} (m-2) \frac{p_{\theta,m-1}}{2^{m-n}}, \label{dmnsum0} \\
d_{mn}^{(1)}
  &\equiv& (-1)^{n+1}\hspace{3pt} _{m-1}{\rm C}_{2} \hspace{3pt} p_{\theta,m-1}, \label{dmnsum1} \\
e_{mn}
  &\equiv& (-1)^{n+1}\hspace{3pt} _{2m-n-5}{\rm C}_{m-3} \hspace{3pt} _{m-3}{\rm C}_{2} \frac{p_{\theta,m-2}}{2^{m-n}}. \label{emnsum} 
\end{eqnarray}
Here
\begin{eqnarray}
p_{\theta,i_r}&=&\frac{(2i_r)!}{2^{2i_r-1}(i_r!)^2} \hspace{10pt} (\text{$i_r$ is a positive integer}), \label{pthetair} \\
C_{f,m}&=&\frac{1}{2}m^2-\frac{5}{8}m,
\end{eqnarray}
where $p_{\theta, i_r}$ is associated with the coefficient of $\cos(i_r\theta)$ in the Legendre polynomial $P_{i_r}(\cos \theta)$ 
(see Appendix~\ref{section_e}). Then, we have $c_{32}=-45/16$ and $d_{32}= -3/8$ for the $3:2$ resonance, $c_{31}=265/64$ and $d_{31}= 9/8$ for the $3:1$ resonance, and $c_{21}=9/4$ for the $2:1$ resonance. As a result of the approximation ignoring the higher order in $\mu, e_J, a, e$,  Eq.~(\ref{myuvemn}) does not include the order $\mu e_J^2$ in the first-order resonance, while Eq.~(\ref{myuvemn}) includes the order $\mu e_J^2$ in the second-order resonance.

In addition, we need the angle-independent terms $\mu V_{{\rm e},0,0,0}'$ in Eq.~(\ref{eq_ell-three-bodies-hamiltonian_fj_sepa_lg_muvep2}). For this calculation we note that the expansion in $e$ as $\cos f= \cos l+e(\cos{2l}-1)+O(e^2)$ in Eq.~(\ref{eq_cosf-el}) contains angle-independent term in the first-order term of $e$ in $\cos f$.
As a result, not only the terms which do not include $f,\theta, f_J$ but also those terms become the angle-independent secular terms. At each of the 3:1, the 2:1 and the 3:2 resonances, the angle-independent secular term is given by
\begin{equation}
\mu V_{{\rm e},0,0,0}'(L,G;\mu)
  \approx 
  -\mu +\mu a^{-1} -\frac{1}{4}\mu a^2\Big(1+\frac{3}{2}e^2\Big),
\label{muvp000mn}
\end{equation}
where the first term $-\mu$ does not contribute to the equation of motion, since this is just a constant.

In the following sections we will use the equal sign ``$=$" instead of the approximation sign ``$\approx$" to avoid complicated expressions.

\section{Single-resonance Hamiltonian  vs. Double-resonance Hamiltonian}
\label{reduction}

This section is the main part of this paper. We will show that our approximated Hamiltonian for the 3:2 resonance reduces to a single-resonance Hamiltonian which is integrable, while approximated Hamiltonians for the 3:1 and the 2:1 resonances
reduce to double-resonance Hamiltonians which are nonintegrable.

In order to see this we first perform a canonical transformation to the approximated Hamiltonian around the $m:n$ resonance point.
 As we will see,  the degree of freedom of the Hamiltonian can be reduced by one degree of freedom by the canonical transformation.

Our approximated Hamiltonian has the following from:
\begin{eqnarray}
H_{f_J}'(L,G,l,g,f_J)
  &=& H'_0 +\mu V_{{\rm e},0,0,0}' +\mu V_{{\rm e},m,n}',
\label{eq_ell_mnresonance-approximation-hamiltonian}
\end{eqnarray}
with Eqs.~(\ref{eq_ell-three-bodies-hamiltonian_fj_sepa_lg_non-perturbation}), (\ref{myuvemn}) and (\ref{muvp000mn}),
where $\mu e_J^2 E_{mn}'(L,G)$ in  Eq.~(\ref{myuvemn}) vanishes in the first-order resonance. 

This $f_J$-dependent Hamiltonian with the time $f_J$ has two degrees of freedom. 
Note that it depends only on the two angles, $mg+nl$ and $g+f_J$, instead of $g$, $l$, and $f_J$ separately. This implies that we can reduce one degree of freedom of the Hamiltonian by a suitable canonical transformation\cite{t}.

The canonical transformation from Delaunay variables $(L,G,l,g)$ to new variables $(y_1,y_2,\phi_1,\phi_2)$ is given by
\begin{eqnarray}
{y_1}&=& \frac{1}{n} L-\tilde{y}_{mn} = \frac{1}{n}(\sqrt{a}-\sqrt{a_{mn}}),
\label{eq_y1} \\
{y_2}&=& G-\frac{m}{n} L, 
\label{eq_y2} \\
{\phi_1}&=& 
  \begin{cases}
   mg+nl & (c_{mn}>0) \\
   mg+nl-\pi & (c_{mn}<0),
  \end{cases}
\label{eq_phi1} \\
{\phi_2}&=& g+f_J -\pi ,
\label{eq_phi2}
\end{eqnarray}
where $\tilde{y}_{mn} \equiv \sqrt{a_{mn}}/n$ (see Appendix \ref{section_f}). The inverse transformations on the canonical momenta are given by
\begin{eqnarray}
L&=&n (y_1+\tilde{y}_{mn}),
\label{eq_L-y} \\
G&=&y_2+m (y_1+\tilde{y}_{mn}).
\label{eq_G-y}
\end{eqnarray}
This canonical transformation leads to an $f_J$-independent Hamiltonian with two degrees of freedom,
\begin{equation}
H
  = H_{f_J}' +y_2 ,
\label{eq_newH}
\end{equation}
which is written by the new variables as
\begin{equation}
H(y_1,y_2,\phi_1,\phi_2)
  =H''_0 +\mu V_{{\rm e},0,0,0}'' +\mu V_{{\rm e},m,n}'',
\label{eq_ell_hamiltonian_res_yphi}
\end{equation} 
with
\begin{eqnarray}
H''_0 (y_1,y_2)
  &=& H'_0 + y_2,  
\label{eqH0y1y2} \\
\mu V_{{\rm e},0,0,0}''({y_1},{y_2})
  &=&\mu V_{{\rm e},0,0,0}'(L,G),
\label{eqve000y1y2}
\end{eqnarray}
where $H_0'$ is given by Eq.~(\ref{eq_ell-three-bodies-hamiltonian_fj_sepa_lg_non-perturbation}), and
\begin{eqnarray}
\mu V_{{\rm e},m,n}''(y_1.y_2,\phi_1,\phi_2) \hspace{-50pt} \nonumber \\
    &=& \mu C_{mn}''(y_1.y_2) \cos \phi_1 \nonumber \\
    & &+\mu e_J D_{mn}''(y_1.y_2) \cos [\phi_1-\phi_2] \nonumber \\
    & &+\mu e_J^2 E_{mn}''(y_1.y_2) \cos [\phi_1- 2\phi_2],
\label{myuvemny1y2}
\end{eqnarray} 
where
\begin{eqnarray}
C_{mn}''({y_1},{y_2})
  &=&
  \begin{cases}
   C_{mn}'(L,G) & (c_{mn}>0) \\
   -C_{mn}'(L,G)  & (c_{mn}<0),
  \end{cases} \label{eq_cmnpp} \\
D_{mn}''({y_1},{y_2})
  &=&
  \begin{cases}
   -D_{mn}'(L,G)  & (c_{mn}>0) \\
   D_{mn}'(L,G)  & (c_{mn}<0),
  \end{cases} \label{eq_dmnpp}\\
  E_{mn}''({y_1},{y_2})
  &=&
  \begin{cases}
   E_{mn}'(L,G) & (c_{mn}>0) \\
   -E_{mn}'(L,G) & (c_{mn}<0).
  \end{cases} \label{eq_emnpp} 
\end{eqnarray}
Here $L$ and $G$ are functions of $y_1$ and $y_2$(see Eqs.~(\ref{eq_L-y}) and (\ref{eq_G-y})).

The Hamiltonian~(\ref{eq_ell_hamiltonian_res_yphi}) is the so-called triple-resonance Hamiltonian, since it depends on three trigonometric functions, $\cos {\phi_1}$, $\cos [{\phi_1}-{\phi_2}]$, and $\cos [{\phi_1}-2{\phi_2}]$. Multiple-resonance Hamiltonian is generally nonintegrable, while the single-resonance Hamiltonian that consists 
only of a single trigonometric function is integrable in spite of the fact that the single-resonance Hamiltonian has a resonance interaction\cite{l}.

Here we come to the crucial part of this paper. The actual observation of the asteroids shows that the Hilda asteroids around the 3:2 resonance are stable since there are asteroids around the resonance point, while the asteroids around the 2:1 and the 3:1 resonance points are unstable since there are Kirkwood gaps.  This suggests that our Hamiltonian~(\ref{eq_ell_hamiltonian_res_yphi}) might be well-approximated by a single-resonance Hamiltonian for the 3:2 resonances, while it cannot be approximated by a single-resonance Hamiltonian for the 2:1 or the 3:1 resonance\cite{x}.

\begin{figure}[t]
  \begin{center}
\includegraphics[width=8.0cm,height=5cm,keepaspectratio]{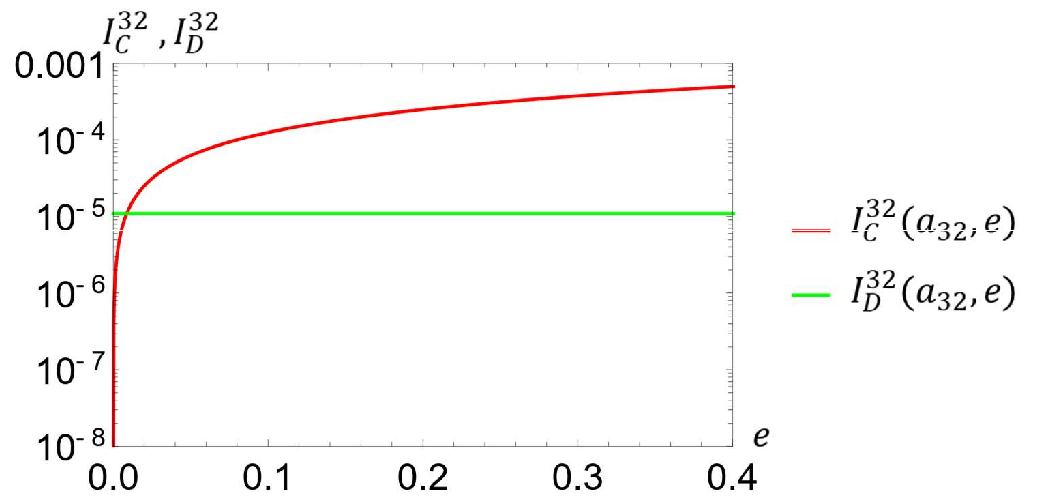}
\caption{The intensities $I_C^{32}(a_{32},e)$ and $I_D^{32}(a_{32},e)$ vs. the eccentricity $e$}
\label{fig_resamp32}
  \end{center}
\end{figure}
\begin{figure}[t]
  \begin{center}
\includegraphics[width=8.0cm,height=5cm,keepaspectratio]{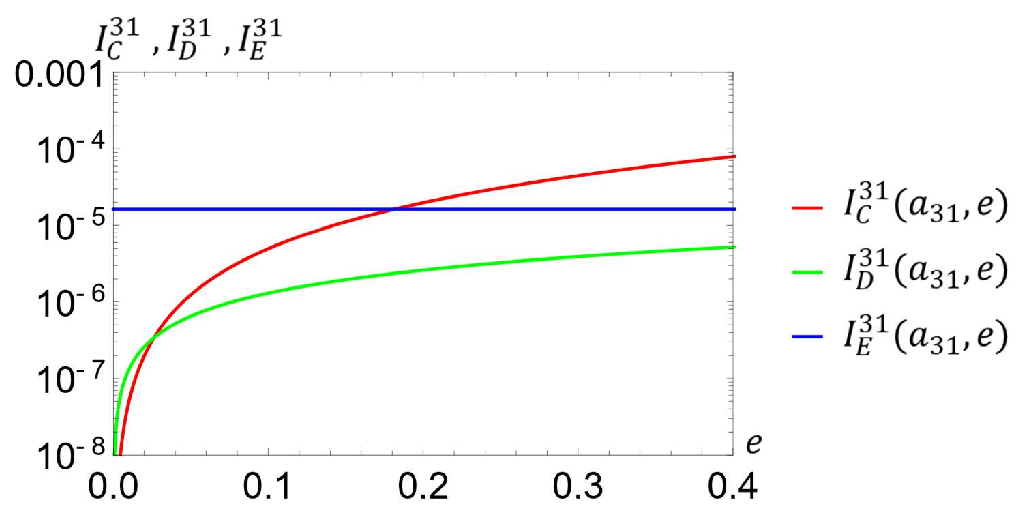}
\caption{The intensities $I_C^{31}(a_{31},e)$, $I_D^{31}(a_{31},e)$ and $I_E^{31}(a_{31},e)$ vs. the eccentricity $e$}
\label{fig_resamp31}
  \end{center}
\end{figure}
\begin{figure}[t]
  \begin{center}
\includegraphics[width=8.0cm,height=5cm,keepaspectratio]{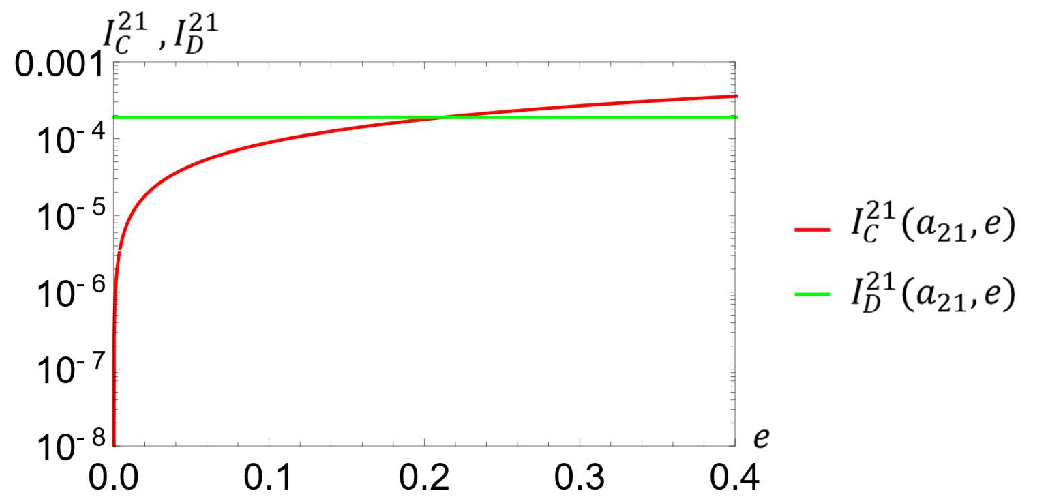}
\caption{The intensities $I_C^{21}(a_{21},e)$ and $I_D^{21}(a_{21},e)$ vs. the eccentricity $e$}
\label{fig_resamp21}
  \end{center}
\end{figure}

Indeed, one can verify this conjecture in the following method:
Let us introduce the quantities that indicate the intensity of each coefficient in Eq.~(\ref{myuvemny1y2}), $\mu C_{mn}''({y_1},{y_2})$,  $\mu e_J D_{mn}''({y_1},{y_2})$, and $\mu e_J^2 E_{mn}''({y_1},{y_2})$, as
\begin{eqnarray}
I_C^{mn}(a,e)
   &\equiv &\abs{\mu C_{mn}''({y_1},{y_2})}=\abs{ \mu c_{mn}a^m e^{m-n} }, \label{eq_fcmn} \\ 
I_D^{mn}(a,e)
    &\equiv &\abs{\mu e_J D_{mn}''({y_1},{y_2})} \nonumber \\
   &=&
  \begin{cases}
    \frac{3}{2}\mu e_J a^{-2} &\hspace{-60pt} \text{for the 2:1 resonance case}\\
    \abs{ \mu e_J d_{mn}a^{m-1} e^{m-n-1} } \\ &\hspace{-60pt} \text{for the other resonance case,}
   \end{cases}
   \label{eq_fdmn} \\
I_E^{mn}(a,e)
    &\equiv& \abs{\mu e_J^2 E_{mn}''({y_1},{y_2})} \nonumber \\
    &=&
    \begin{cases}
    0 &\hspace{-60pt} \text{for the first-order resonance} \\
    \frac{3}{2}\mu e_J^2a^{-2} &\hspace{-60pt} \text{for the 3:1 resonance case} \\
    \abs{ \mu e_J^2 e_{mn}a^{m-2} e^{m-n-2} } \\ &\hspace{-60pt} \text{for the other second-order resonance.}
     \end{cases}
     \label{eq_femn}
\end{eqnarray}
We show these intensities as a function of the eccentricity $e$ of the asteroid for the 3:2 resonance in Fig.~\ref{fig_resamp32}, for the 3:1 resonance in Fig.~\ref{fig_resamp31}, and for the 2:1 resonance in Fig.~\ref{fig_resamp21}, respectively. As mentioned just after Eq.~(\ref{eq_ell_mnresonance-approximation-hamiltonian}), $I_E^{mn}$ is negligible for the first-order resonances (with the 3:2 and the 2:1 resonances).

From Fig.~\ref{fig_resamp32} one can see for the 3:2 resonance that $I_D^{32}(a,e)$ is much smaller than $I_C^{32}(a,e)$ except extremely small eccentricity with $e<0.02$.  The numerical calculation of the equations of motion with the Hamiltonian~(\ref{eq_ell-three-bodies-hamiltonian_fj}) starting with the initial condition of the eccentricity $e=0$ shows the value of $e$ rapidly reaches about $e=0.1$ which is much larger value than $e=0.02$.
Then $I_C^{32}(a,e)$ is much larger than $I_D^{32}(a,e)$. The Hamiltonian~(\ref{eq_ell_hamiltonian_res_yphi}) is well-approximated by a single-resonance Hamiltonian with $ \cos {\phi_1}$ term for the 3:2 resonance.
Hence, the system is integrable, and the motion is regular for the 3:2 resonance.

On the other hand, from Fig.~\ref{fig_resamp31} one can see for the 3:1 resonance that $I_E^{31}(a,e)$ has more or less the same intensity as $I_C^{31}(a,e)$ around the typical value of the eccentricity $e\approx 0.2$ ($0.1<e<0.3$). The Hamiltonian~(\ref{eq_ell_hamiltonian_res_yphi}) in this 3:1 resonance case is a typical double-resonance Hamiltonian. Hence, the system is nonintegrable, and the motion is chaotic for the 3:1 resonance.

From Fig.~\ref{fig_resamp21} one can see for the 2:1 resonance that  $I_D^{21}(a,e)$ has more or less the same intensity as $I_C^{21}(a,e)$.  
The Hamiltonian~(\ref{eq_ell_hamiltonian_res_yphi}) in this case is also a typical double-resonance Hamiltonian.  Hence,  the system is nonintegrable, and the motion is chaotic for the 2:1 resonance around $0.1<e<0.3$.

These results show the reason why there are stable Hilda asteroids around the 3:2 resonance in spite of the fact that there is a resonance interaction, while there are Kirkwood gaps around the 2:1 and the 3:1 resonances due to the resonance interaction.


\section{Poincar\'e surfaces of section}
\label{poincare32}

\begin{figure*}[bthp]
  \begin{center}
    \begin{tabular}{l}
      \begin{minipage}{0.50\hsize}
        \begin{center}
          \includegraphics[height=3.2cm,keepaspectratio]{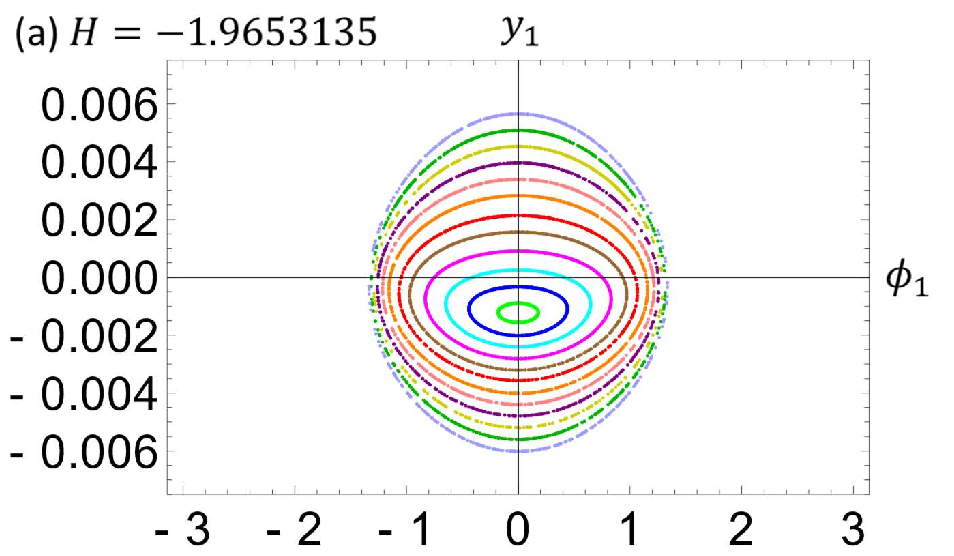}
        \end{center}
      \end{minipage}
      \begin{minipage}{0.50\hsize}
        \begin{center}
          \includegraphics[height=3.2cm,keepaspectratio]{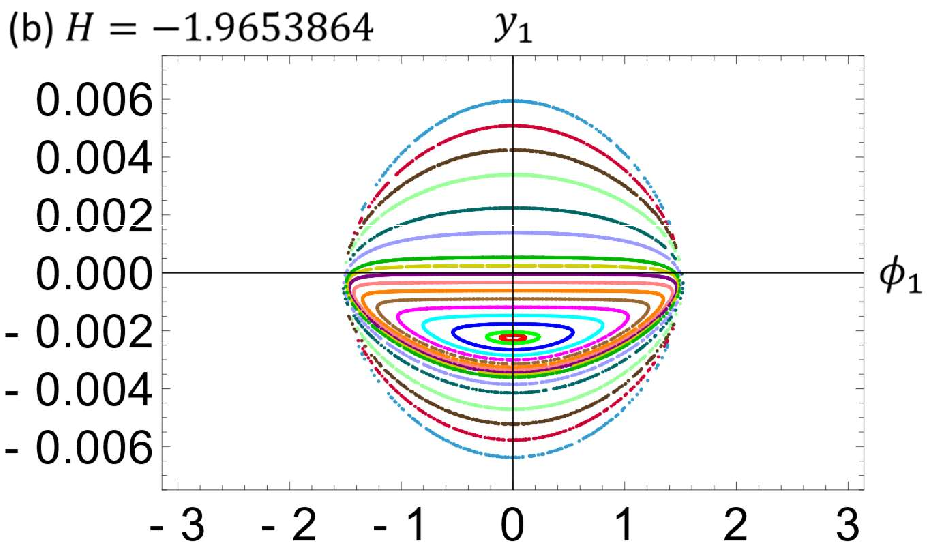}
        \end{center}
      \end{minipage}\\ \\
      \begin{minipage}{0.50\hsize}
        \begin{center}
          \includegraphics[height=3.2cm,keepaspectratio]{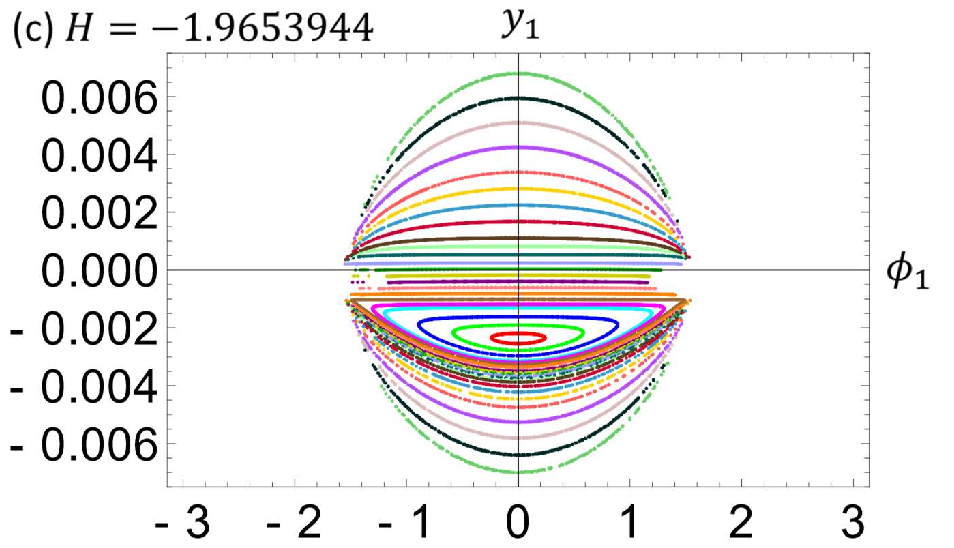}
        \end{center}
      \end{minipage}
       \begin{minipage}{0.50\hsize}
        \begin{center}
          \includegraphics[height=3.2cm,keepaspectratio]{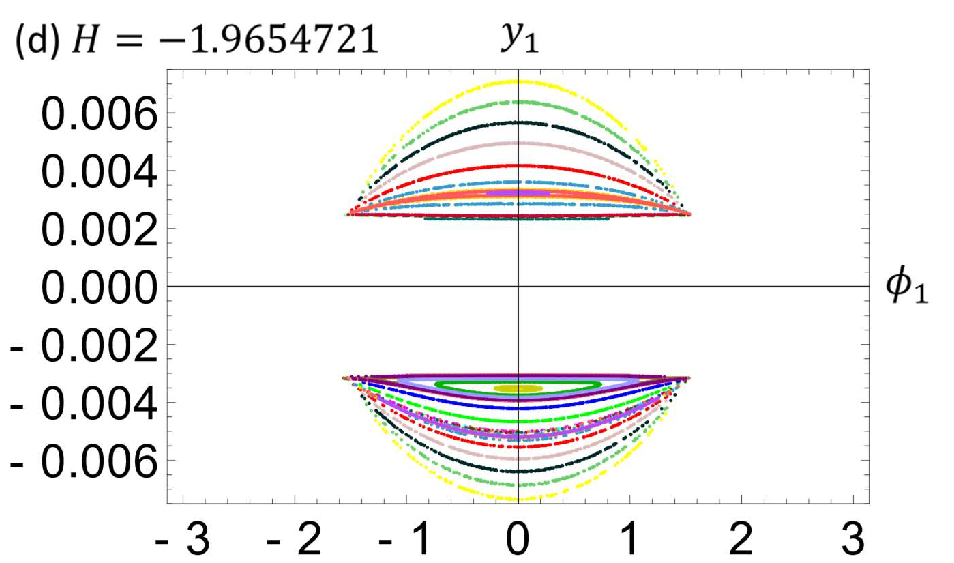}
        \end{center}
      \end{minipage}
    \end{tabular}
    \caption{Poincar\'e surfaces of section $(\phi_1,y_1)$ around the 3:2 resonance point. (a) $H=-1.9653135$ with the initial conditions $(a, e, l, g, f_J)=(a_{32}, 0.14, \pi/3, 0, \pi)$. (b) $H=-1.9653864$ with the initial conditions $(a, e, l, g, f_J)=(0.7715, 0.14, \pi/3, 0, \pi)$. (c) $H=-1.9653944$  with the initial conditions $(a, e, l, g, f_J)=(0.7720, 0.14, \pi/3, 0, \pi)$. (d) $H=-1.9654721$ with the initial conditions $(a, e, l, g, f_J)=(a_{32}, 0.14, 2\pi/3, \pi/3, \pi)$}
    \label{fig_32respsphi1y1}
  \end{center}
\end{figure*}
\begin{figure*}[pbht]
  \begin{center}
    \begin{tabular}{l}
        \begin{minipage}{0.50\hsize}
        \begin{center}
          \includegraphics[height=3.2cm,keepaspectratio]{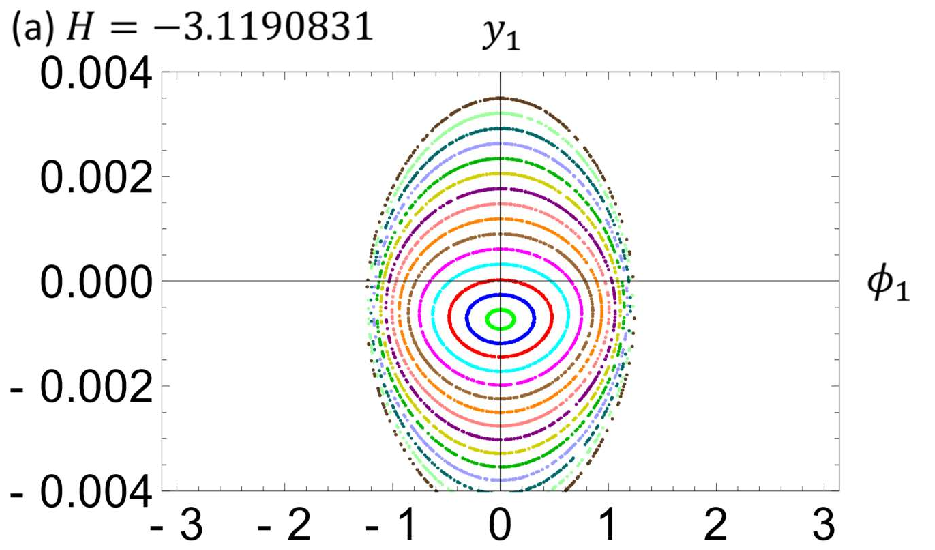}
        \end{center}
      \end{minipage}
        \begin{minipage}{0.50\hsize}
        \begin{center}
          \includegraphics[height=3.2cm,keepaspectratio]{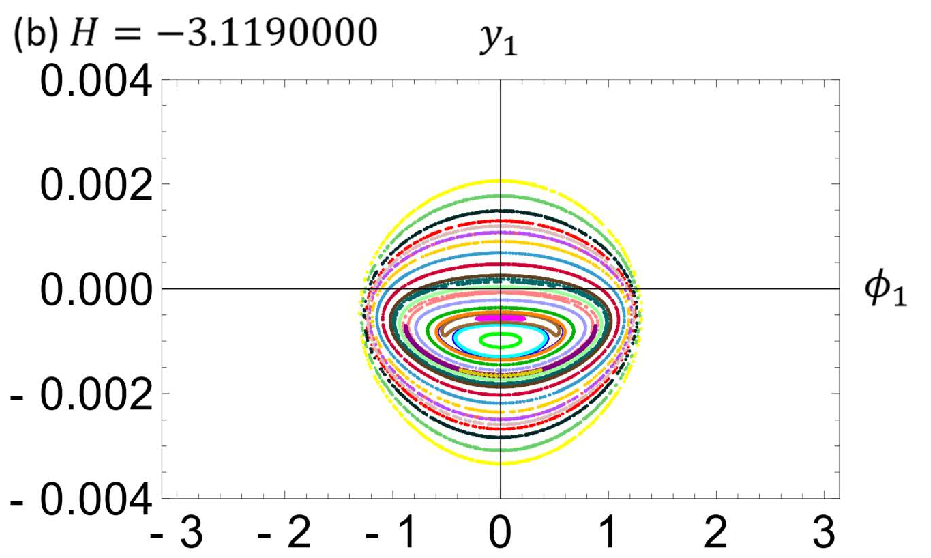}
        \end{center}
      \end{minipage}\\ \\
      \begin{minipage}{0.50\hsize}
        \begin{center}
          \includegraphics[height=3.2cm,keepaspectratio]{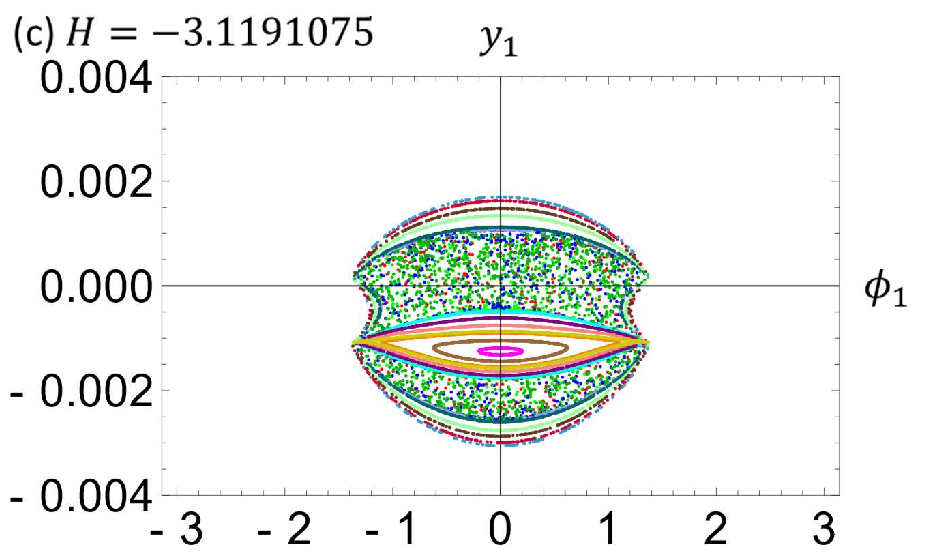}
        \end{center}
      \end{minipage}
       \begin{minipage}{0.50\hsize}
        \begin{center}
          \includegraphics[height=3.2cm,keepaspectratio]{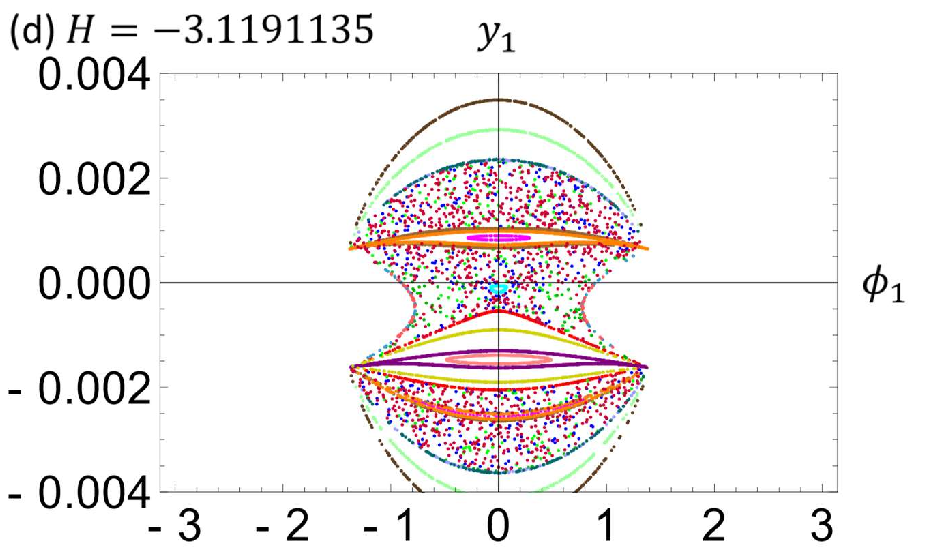}
        \end{center}
      \end{minipage}
    \end{tabular}
    \caption{Poincar\'e surfaces of section $(\phi_1,y_1)$ around the 3:1 resonance point. (a) $H=-3.1190831$ with the initial conditions $(a, e, l, g, f_J)=(a_{31}, 0.14, 2\pi/3, \pi/3, \pi)$. (b) $H=-3.1191000$ with the initial conditions $(a, e, l, g, f_J)=(a_{31}, 0.14, \pi/3, \pi/3, \pi)$. (c) $H=-3.1191075$ with the initial conditions $(a, e, l, g, f_J)=(a_{31}, 0.14, 2\pi/3, \pi, \pi)$. (d) $H=-3.1191135$ with the initial conditions $(a, e, l, g, f_J)=(a_{31}, 0.14, 0, 0, \pi)$}
    \label{fig_31respsphi1y1}
  \end{center}
\end{figure*}
\begin{figure*}[pbht]
  \begin{center}
    \begin{tabular}{l}
        \begin{minipage}{0.50\hsize}
        \begin{center}
          \includegraphics[height=3.2cm,keepaspectratio]{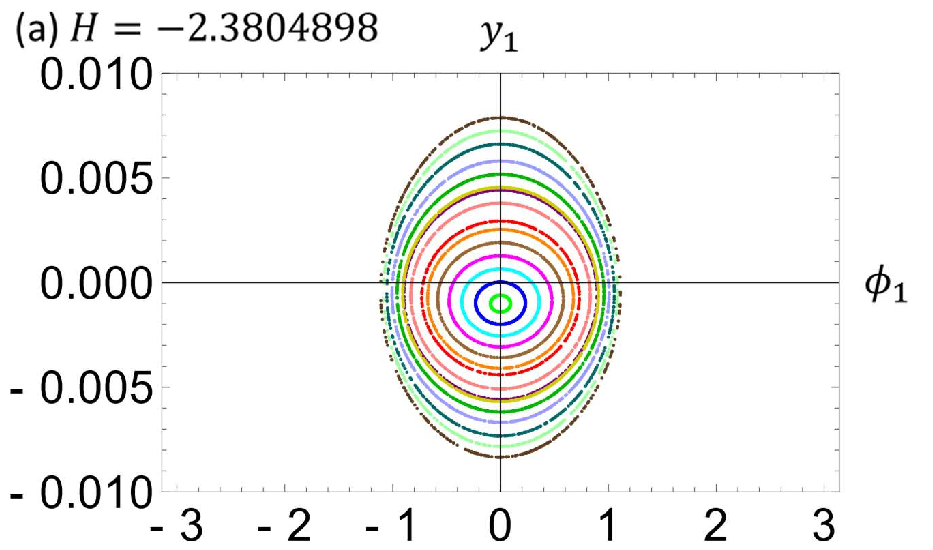}
        \end{center}
      \end{minipage}
        \begin{minipage}{0.50\hsize}
        \begin{center}
          \includegraphics[height=3.2cm,keepaspectratio]{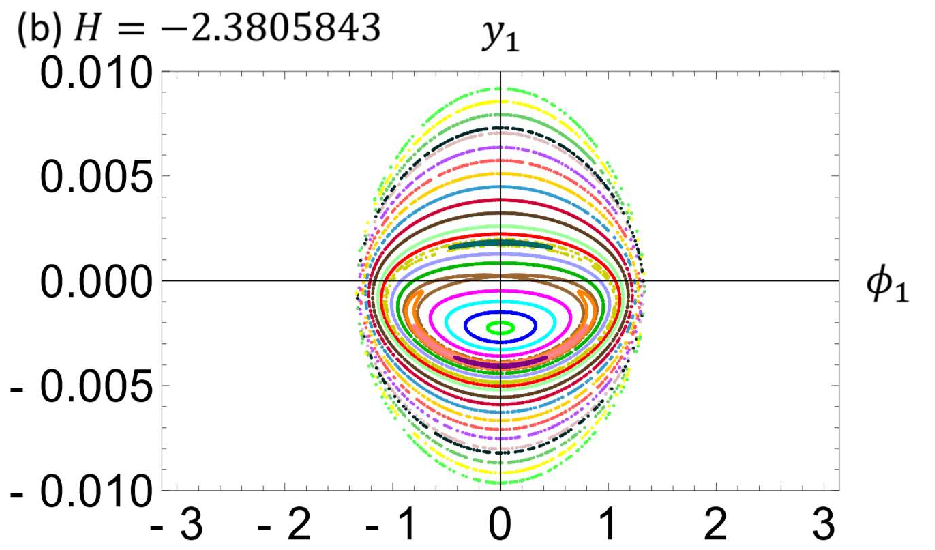}
        \end{center}
      \end{minipage}\\ \\
      \begin{minipage}{0.50\hsize}
        \begin{center}
          \includegraphics[height=3.2cm,keepaspectratio]{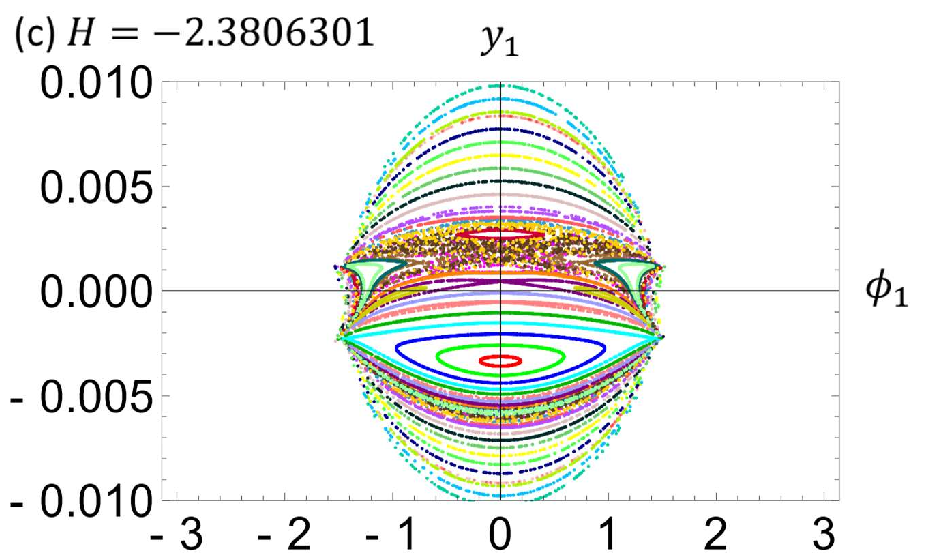}
        \end{center}
      \end{minipage}
       \begin{minipage}{0.50\hsize}
        \begin{center}
          \includegraphics[height=3.2cm,keepaspectratio]{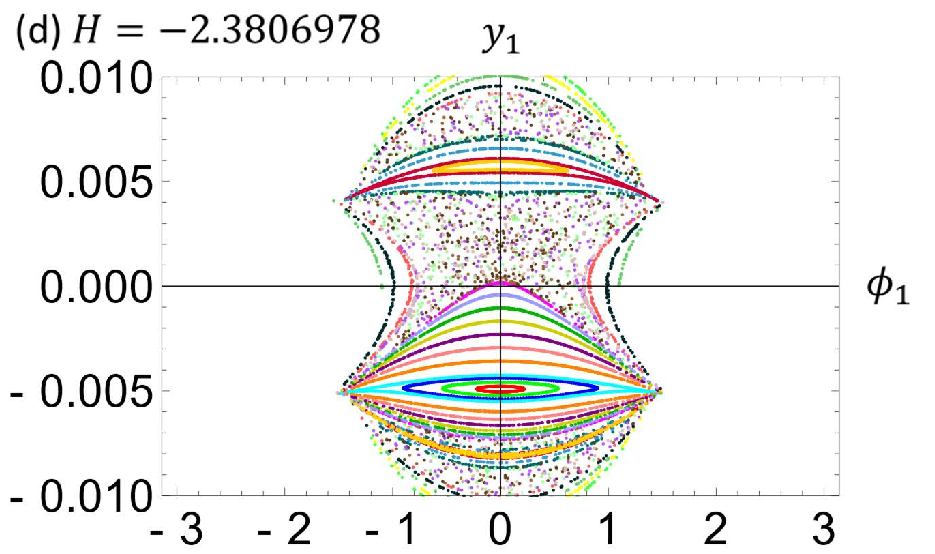}
        \end{center}
      \end{minipage}
    \end{tabular}
    \caption{Poincar\'e surfaces of section $(\phi_1,y_1)$ around the 2:1 resonance point. (a) $H=-2.3804898$ with the initial conditions $(a, e, l, g, f_J)=(a_{21}, 0.14, 2\pi/3, \pi, \pi)$. (b) $H=-2.3805843$ with the initial conditions $(a, e, l, g, f_J)=(a_{21}, 0.14, 2\pi/3, 0, \pi)$. (c) $H=-2.3806301$ with the initial conditions $(a, e, l, g, f_J)=(0.6345, 0.14, 2\pi/3, 0, \pi)$. (d) $H=-2.3806978$ with the initial conditions $(a, e, l, g, f_J)=(0.6350, 0.14, \pi, 0, \pi)$}
    \label{fig_21respsphi1y1}
  \end{center}
\end{figure*}

In this section, we  construct Poincar\'e surfaces of section
\cite{l, o} for the Hamiltonian~(\ref{eq_ell_hamiltonian_res_yphi}).  This is possible because the motion is generated by a time-independent Hamiltonian with two degrees of freedom. By plotting the Poincar\'e surfaces of section, we can verify the prediction on the regular or chaotic motion of the asteroid discussed in the previous section.

We calculate trajectories of asteroids in the phase space $(y_1,y_2,\phi_1,\phi_2)$ 
 by numerically solving the canonical equations of motion generated by the Hamiltonian~(\ref{eq_ell_hamiltonian_res_yphi}).
We calculate the trajectories with several initial conditions for some energies(i.e., values of the Hamiltonian) of the system at each resonance. We construct the intersection of the trajectories in a two-dimensional surface $(\phi_1,y_1)$ given by $\phi_2=2\pi \bar{n}(\bar{n}=...,-2,-1,0,1,2,...)$, where $\phi_2=g+f_J -\pi$ is the longitude of the apsidal line of the asteroid relative to the fixed system.   We  construct the Poincar\'e surfaces of section for $\partial \phi_2 / \partial f_J < 0$. We note that the resonance point at $a=a_{mn}$ corresponds to the point $y_1 =0$ (see Eq.~(\ref{eq_y1})).

Poincar\'e surfaces of section for $(\phi_1,y_1)$ for the 3:2 resonance are presented for several energies of the system in Fig.~\ref{fig_32respsphi1y1}. Points with different color correspond to different initial conditions of an asteroid.
We have well-defined lines in all surfaces of section in this figure.
As it was predicted in the previous section, the motion is regular around this resonance. Hence we can conclude that the Hamiltonian for the 3:2 resonance is well approximated by the single-resonance Hamiltonian.

Poincar\'e surfaces of section for $(\phi_1,y_1)$ for the 3:1 resonance are presented 
in Fig.~\ref{fig_31respsphi1y1}. The chaotic motion clearly appears as scattered points on the surfaces of section shown in Fig.~\ref{fig_31respsphi1y1} (c) and (d). This is consistent with the fact that the Hamiltonian for the 3:1 resonance is approximated by the double-resonance Hamiltonian.
Poincar\'e surfaces of section for $(\phi_1,y_1)$ for the 2:1 resonance are presented in Fig.~\ref{fig_21respsphi1y1}. The chaotic motion appears in this resonance on the surfaces of section shown in Fig.~\ref{fig_21respsphi1y1} (c) and (d).
This is also consistent with the double resonance Hamiltonian for the 2:1 resonance that is approximately obtained in the previous section.

One of the interesting results of these Poincar\'e surfaces of section for the 3:1 and the 2:1 resonances is that one cannot see typical chaotic motions for relatively high energy cases. Indeed, regular motions are observed in Fig.~\ref{fig_31respsphi1y1} (a), (b) and in Fig.~\ref{fig_21respsphi1y1} (a), (b).

\begin{figure*}[thbp]
  \begin{center}
    \begin{tabular}{l}
      \begin{minipage}{0.50\hsize}
        \begin{center}
          \includegraphics[height=3.3cm,keepaspectratio]{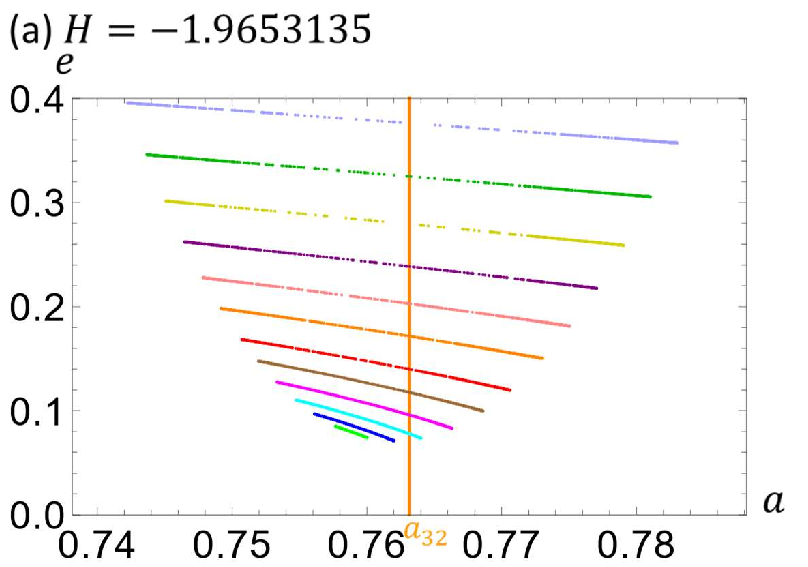}
        \end{center}
      \end{minipage}
      \begin{minipage}{0.50\hsize}
        \begin{center}
          \includegraphics[height=3.3cm,keepaspectratio]{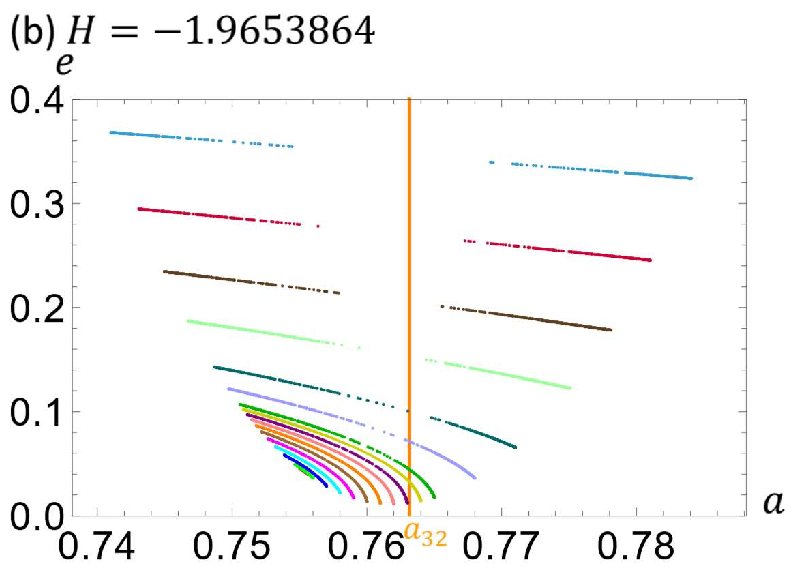}
        \end{center}
      \end{minipage}\\ \\
       \begin{minipage}{0.50\hsize}
        \begin{center}
          \includegraphics[height=3.3cm,keepaspectratio]{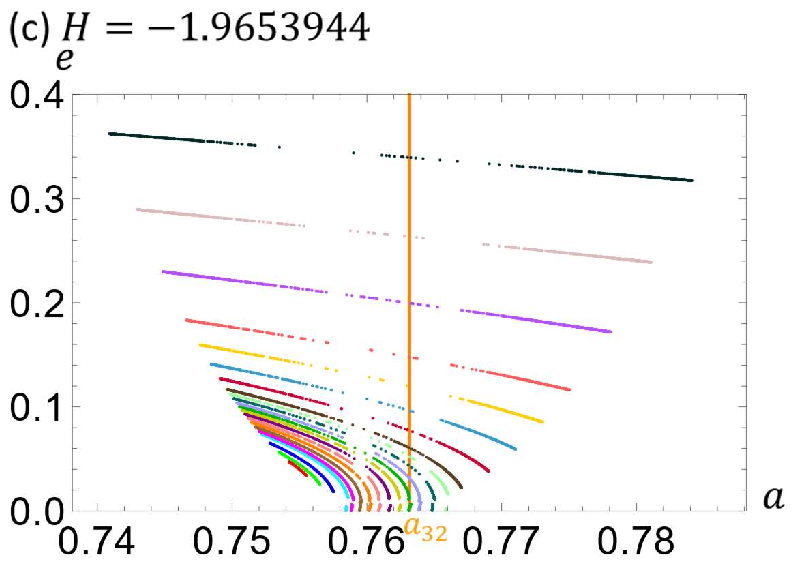}
        \end{center}
      \end{minipage}
      \begin{minipage}{0.50\hsize}
        \begin{center}
          \includegraphics[height=3.3cm,keepaspectratio]{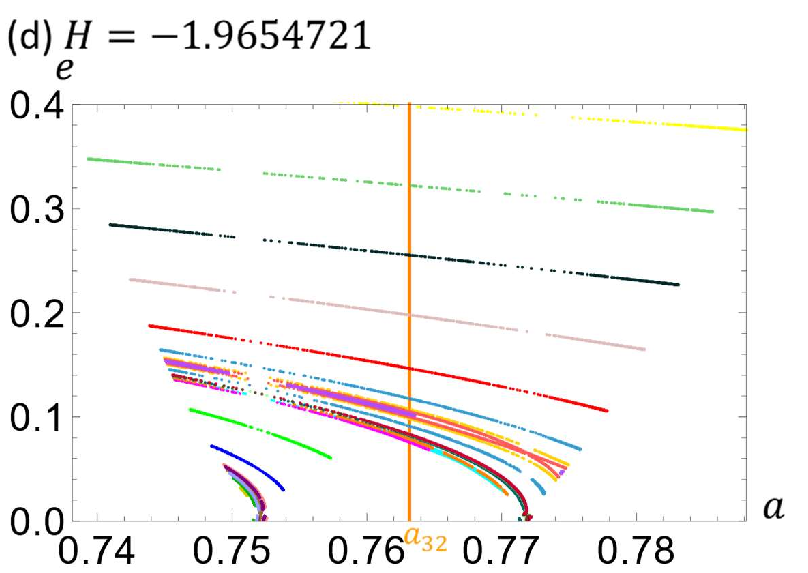}
        \end{center}
      \end{minipage}
    \end{tabular}
    \caption{Surfaces of section $(a,e)$ around the 3:2 resonance point corresponding to Fig.~\ref{fig_32respsphi1y1}. These are not exactly Poincar\'e surfaces of section because the set of the points does not preserve the area. Figs.~(a), (b), (c), (d) are for $H=-1.9653135$, $-1.9653864$, $-1.9653944$, and $-1.9654721$, respectively. Center line of each figure corresponds to the 3:2 resonance point, $a=a_{32}$}
    \label{fig_32respsae}
  \end{center}
\end{figure*}

\begin{figure*}[pbht]
  \begin{center}
    \begin{tabular}{l}
       \begin{minipage}{0.50\hsize}
        \begin{center}
          \includegraphics[height=3.3cm,keepaspectratio]{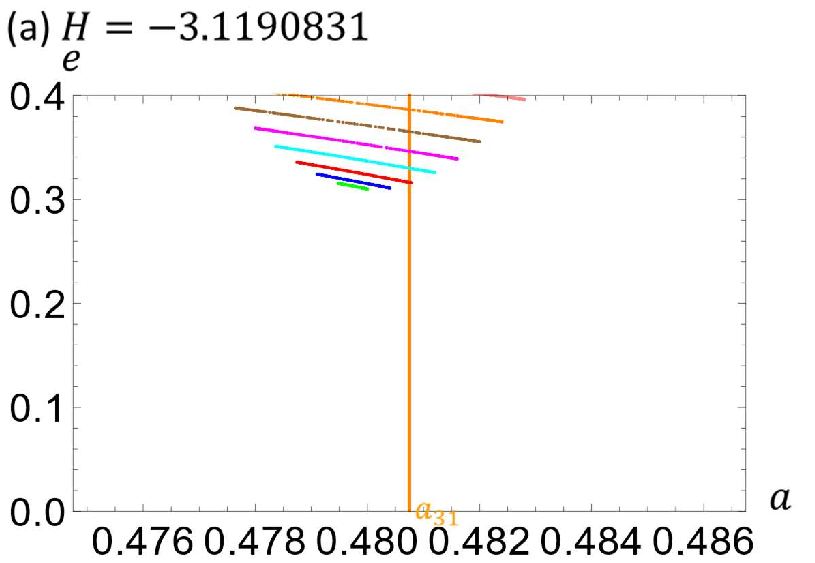}
        \end{center}
      \end{minipage}
      \begin{minipage}{0.50\hsize}
        \begin{center}
          \includegraphics[height=3.3cm,keepaspectratio]{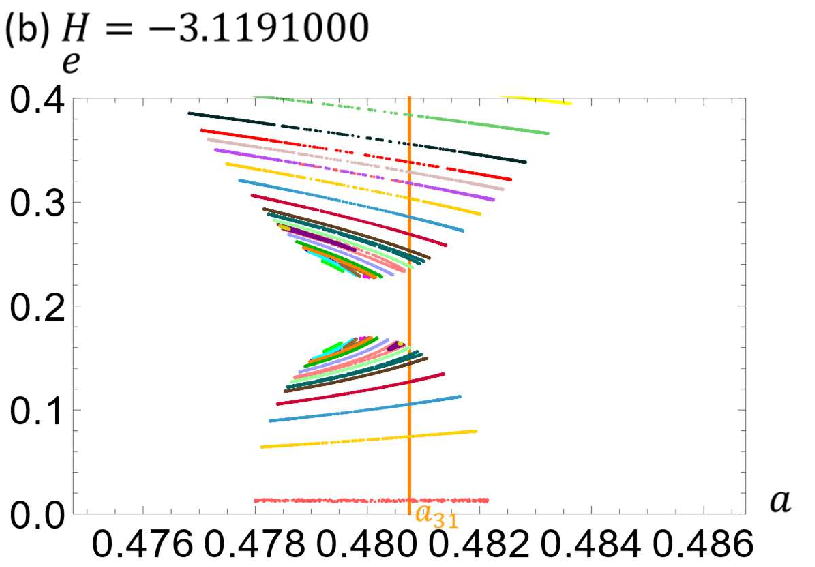}
        \end{center}
      \end{minipage}\\ \\
      \begin{minipage}{0.50\hsize}
        \begin{center}
          \includegraphics[height=3.3cm,keepaspectratio]{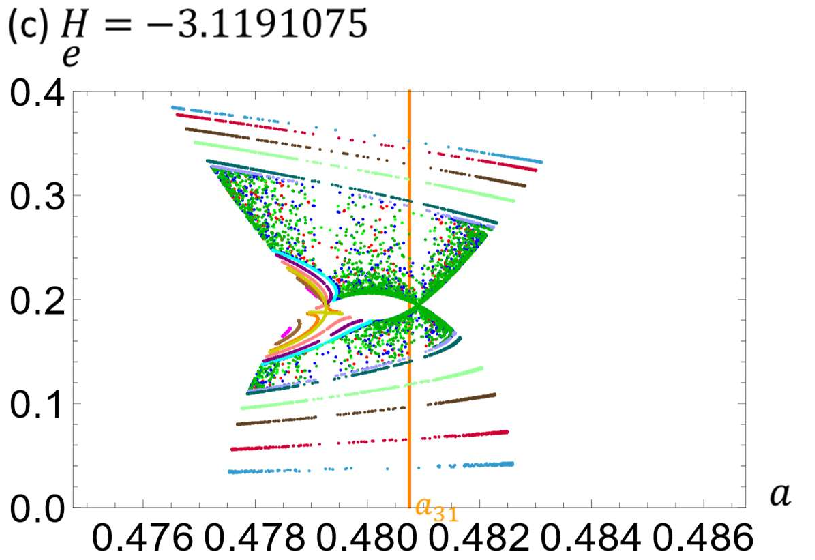}
        \end{center}
      \end{minipage}
      \begin{minipage}{0.50\hsize}
        \begin{center}
          \includegraphics[height=3.3cm,keepaspectratio]{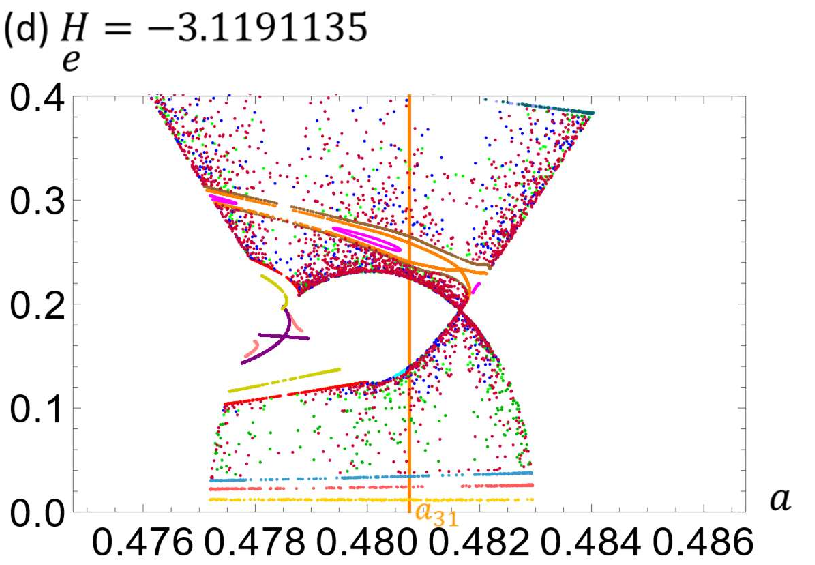}
        \end{center}
      \end{minipage}
    \end{tabular}
    \caption{Surfaces of section $(a,e)$ around the 3:1 resonance point corresponding to Fig.~\ref{fig_31respsphi1y1}. These are not exactly Poincar\'e surfaces of section because the set of the points does not preserve the area. Figs.~(a), (b), (c), (d) are for $H=-3.1190831$, $-3.1191000$,  $-3.1191075$, and $-3.1191135$, respectively. Center line of each figure corresponds to the 3:1 resonance point, $a=a_{31}$}
    \label{fig_31respsae}
  \end{center}
\end{figure*}

\begin{figure*}[pbht]
  \begin{center}
    \begin{tabular}{l}
       \begin{minipage}{0.50\hsize}
        \begin{center}
          \includegraphics[height=3.3cm,keepaspectratio]{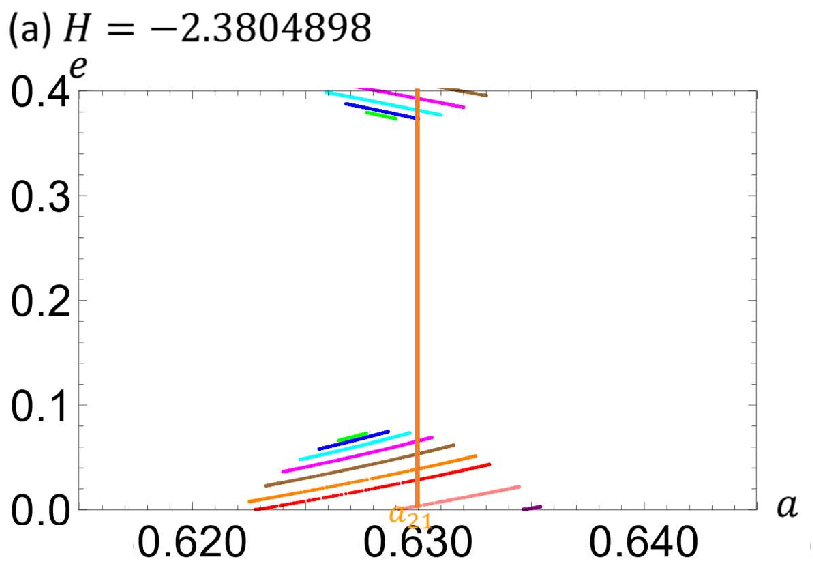}
        \end{center}
      \end{minipage}
      \begin{minipage}{0.50\hsize}
        \begin{center}
          \includegraphics[height=3.3cm,keepaspectratio]{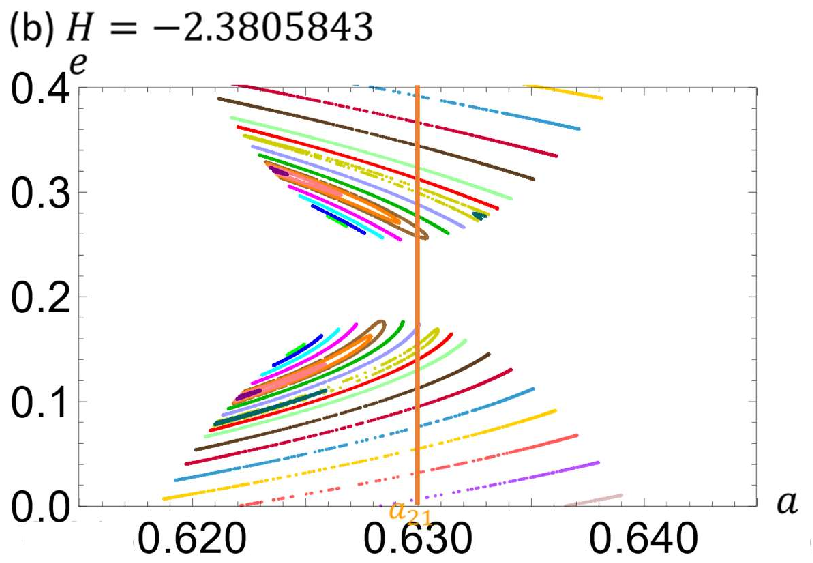}
        \end{center}
      \end{minipage}\\ \\
      \begin{minipage}{0.50\hsize}
        \begin{center}
          \includegraphics[height=3.3cm,keepaspectratio]{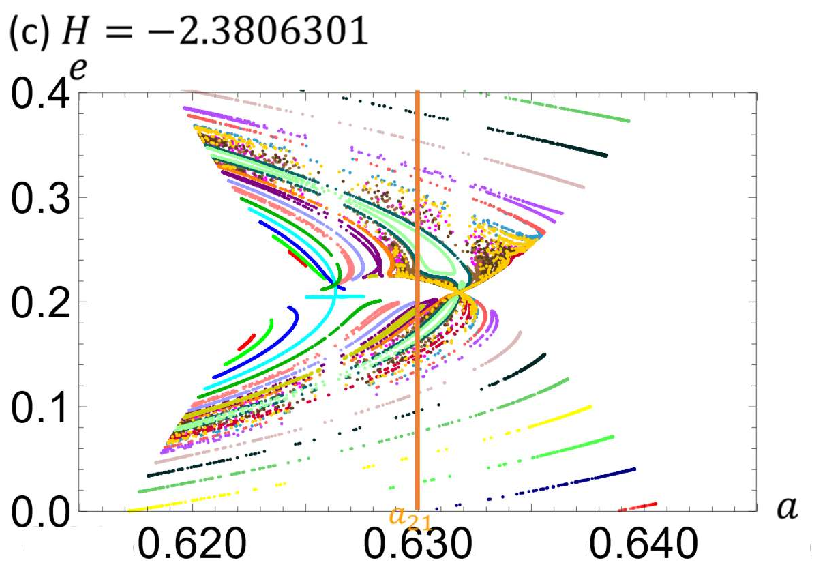}
        \end{center}
      \end{minipage}
      \begin{minipage}{0.50\hsize}
        \begin{center}
          \includegraphics[height=3.3cm,keepaspectratio]{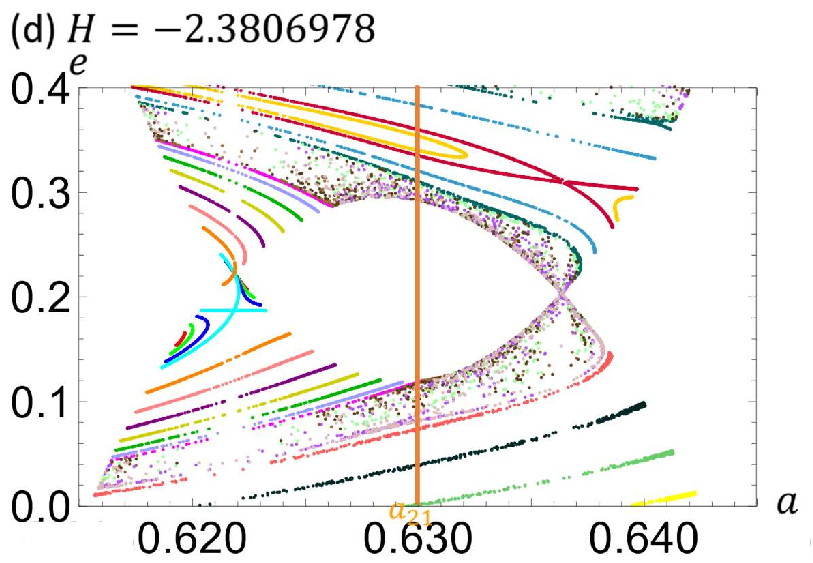}
        \end{center}
      \end{minipage}
    \end{tabular}
    \caption{Surfaces of section $(a,e)$ around the 2:1 resonance point corresponding to Fig.~\ref{fig_21respsphi1y1}. These are not exactly Poincar\'e surfaces of section because the set of the points does not preserve the area. Figs.~(a), (b), (c), (d) are for $H=-2.3804898$, $-2.3805843$, $-2.3806301$, and $-2.3806978$, respectively. Center line of each figure corresponds to the 2:1 resonance point, $a=a_{21}$}
    \label{fig_21respsae}
  \end{center}
\end{figure*}

One can understand this situation by constructing corresponding surfaces of sections for the eccentricity $e$ vs. the semimajor axis $a$ showing in Figs.~\ref{fig_32respsae}, \ref{fig_31respsae}, and \ref{fig_21respsae}. Here each point in these figures  corresponds to the point in Figs.~\ref{fig_32respsphi1y1},  \ref{fig_31respsphi1y1}, and \ref{fig_21respsphi1y1}, respectively.
However, we should note that the surfaces of sections for $(a,e)$ are not exactly Poincar\'e surfaces of section, because the set of the points does not preserve the area\cite{o,p}.

For the 3:2 resonance in Fig.~\ref{fig_32respsae}, one can see a parabola-like shape opening to above in the $a$-$e$ plane.
Each line in the figures corresponds to one given initial condition. 
The regularity of the motion is seen as a regular line in the figure. For all energies shown in Fig.~\ref{fig_32respsae}, the lines are located roughly in the region $0 < e <0.4$.

For the 3:1 resonance, one can see a single parabola-like shape opening to above for the high energy case shown in Fig.~\ref{fig_31respsae} (a). The motion is regular in this case as one can see well-defined lines in the region $e > 0.3$.
Let us compare this figure with Fig.~\ref{fig_resamp31}. For the region $e > 0.3$, $I_C^{31}(a_{31},e)$ is much larger than $I_E^{31}(a_{31})$. Hence, the Hamiltonian~(\ref{eq_ell_hamiltonian_res_yphi}) is, in this region, approximately a single-resonance Hamiltonian that is integrable. 
Next, let us focus our attention on Fig.~\ref{fig_31respsae} (b).
The energy of this system is slightly lower than that of the system in Fig.~\ref{fig_31respsae} (a). The motion is still regular in this case as one can see well-defined lines on the surface. Here we have two parabola-like shapes with the opposite direction. 
Note that, in the distribution of the points along $e$ axis, there is a gap around
 $e = 0.2$ in Fig.~\ref{fig_31respsae} (b).
Since we have no point on the surface of section around $e = 0.2$,
where the intensity of  $I_C^{31}(a_{31},e)$ and $I_E^{31}(a_{31})$ are nearly the same order as  shown in Fig.~\ref{fig_resamp31},
the Hamiltonian~(\ref{eq_ell_hamiltonian_res_yphi}) is still approximately an integrable single-resonance Hamiltonian (see Fig.~\ref{fig_31respsphi1y1}(b)).
For the system shown in Fig.~\ref{fig_31respsae} (c),
the energy of which is slightly lower than that of the system in Fig.~\ref{fig_31respsae} (b), we can see the chaotic motion around $e = 0.2$ 
as scattered points on the surface.
The two parabola-like shapes coalesce around $e = 0.2$,
which is near the intersection point in Fig.~\ref{fig_resamp31}. The Hamiltonian~(\ref{eq_ell_hamiltonian_res_yphi}) is now approximated by a nonintegrable double-resonance Hamiltonian. 
The chaotic region further expands
for the system shown in Fig.~\ref{fig_31respsae} (d), 
the energy of which is the lowest among the systems in Fig.~\ref{fig_31respsae}.

For the 2:1 resonance, we can see the chaotic region around $e =0.2$ 
on the surfaces of section in Figs.~\ref{fig_21respsae} (c) and (d).
One can analyze the surfaces of section shown in Fig.~\ref{fig_21respsae} in the same way as in Fig.~\ref{fig_31respsae} for the 3:1 resonance.

\section{Discussion}
\label{discussion}

In the previous section, we have shown that the motion with the long time average for the 3:2 resonance corresponding to Hilda asteroids is regular, while the motions with long time averages for the 3:1 and the 2:1 resonances corresponding to Kirkwood gaps are chaotic.

%
%

\begin{figure*}[btph]
  \begin{center}
    \begin{tabular}{l}
      \begin{minipage}{0.38\hsize}
        \begin{center}
          \includegraphics[height=4.5cm,keepaspectratio]{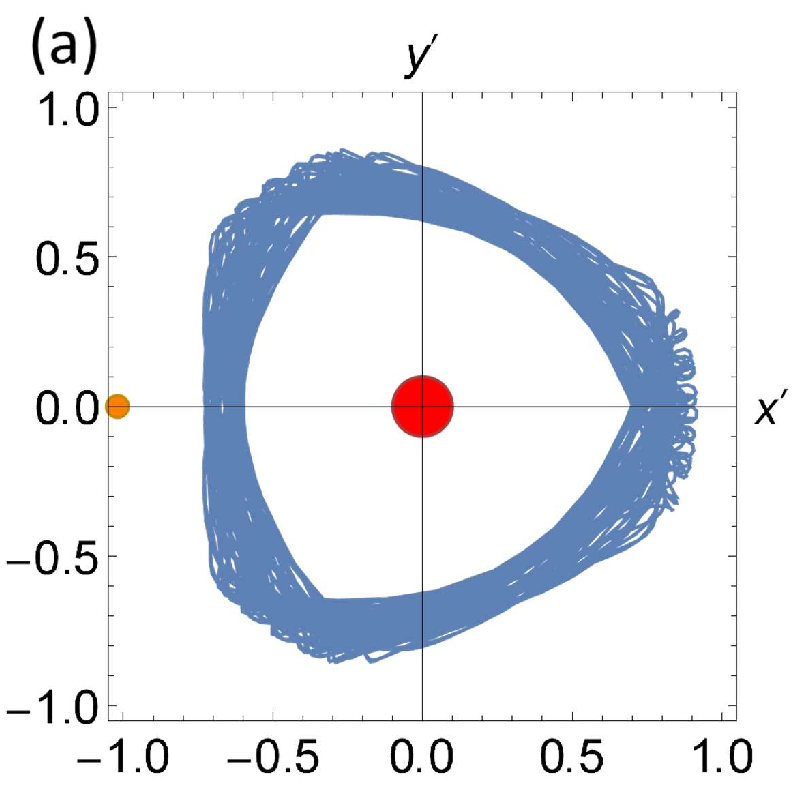}
        \end{center}
      \end{minipage}\hspace{10pt}
      \begin{minipage}{0.38\hsize}
        \begin{center}
          \includegraphics[height=4.5cm,keepaspectratio]{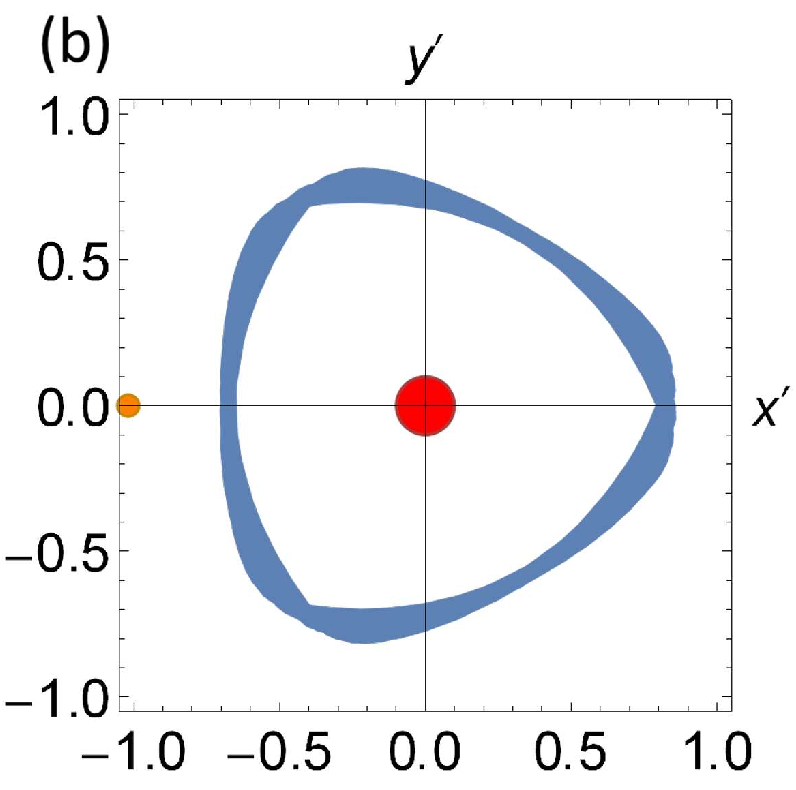}
        \end{center}
      \end{minipage}
    \end{tabular}
    \caption{Numerical results of the trajectories of an asteroid in the rotating system with the initial conditions $(a, e, l, g, f_J)=(a_{32}, 0.10, 2\pi/3, \pi/2, \pi)$. (a) The trajectory is calculated by the original Hamiltonian. (b) The trajectory is calculated by our approximated Hamiltonian. The big circles in the center and the small circles at the left edge are the positions of the sun and Jupiter, respectively. We plot the trajectories for the time $f_J=\pi \sim 800+\pi$. Both trajectories construct Hilda triangle}
  \label{fig_triangle-Hfj}
  \end{center}
\end{figure*}
\begin{figure*}[btph]
  \begin{center}
    \begin{tabular}{l}
      \begin{minipage}{0.45\hsize}
        \begin{center}
          \includegraphics[height=3.0cm,keepaspectratio]{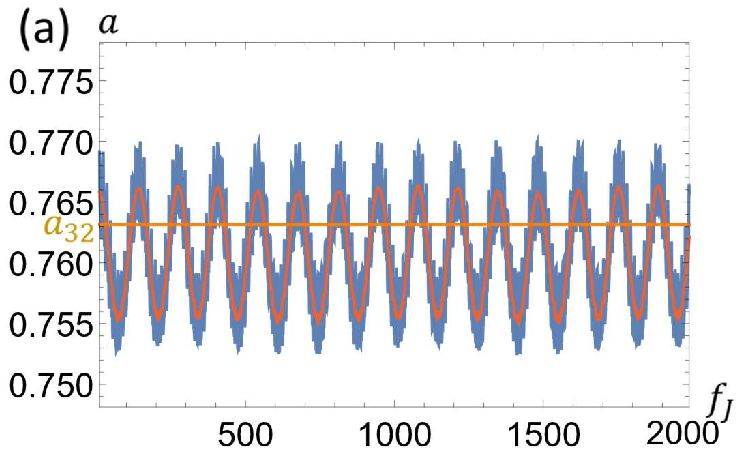}
        \end{center}
      \end{minipage}
      \begin{minipage}{0.45\hsize}
        \begin{center}
          \includegraphics[height=3.0cm,keepaspectratio]{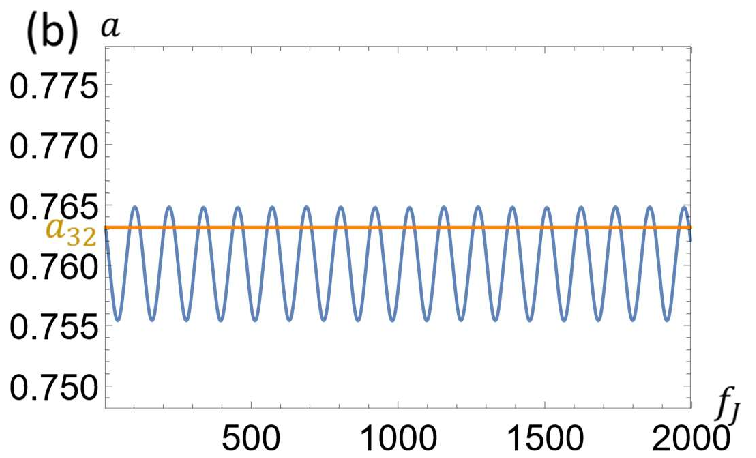}
        \end{center}
      \end{minipage}
    \end{tabular}
    \caption{Numerical results of a time evolution of the semimajor axis $a$ of the asteroid with the time $f_J$ with the initial conditions $(a, e, l, g, f_J)=(a_{32}, 0.10, 2\pi/3, \pi/2, \pi)$. (a) The semimajor axis is calculated by the original Hamiltonian.  (b) The semimajor axis is calculated by our approximated Hamiltonian. Red line in (a) is a time-averaged semimajor axis over $4\pi$ period}
          \label{fig_a-timedependency-Hfj}
  \end{center}
\end{figure*}
\begin{figure*}[btph]
  \begin{center}
    \begin{tabular}{l}
      \begin{minipage}{0.40\hsize}
        \begin{center}
          \includegraphics[height=3.0cm,keepaspectratio]{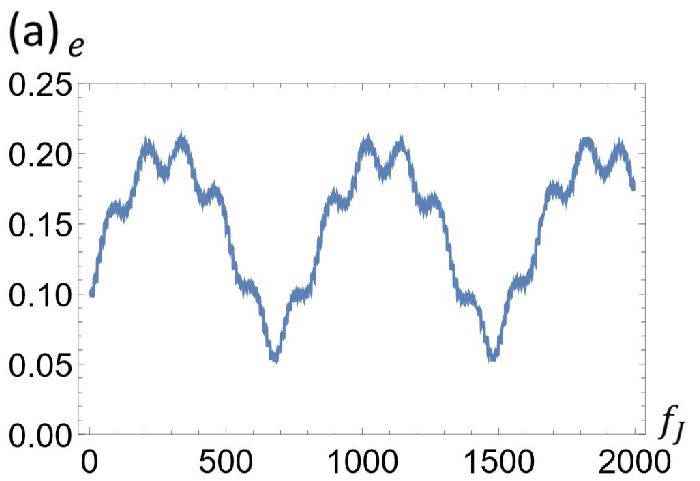}
        \end{center}
      \end{minipage}
      \begin{minipage}{0.40\hsize}
        \begin{center}
          \includegraphics[height=3.0cm,keepaspectratio]{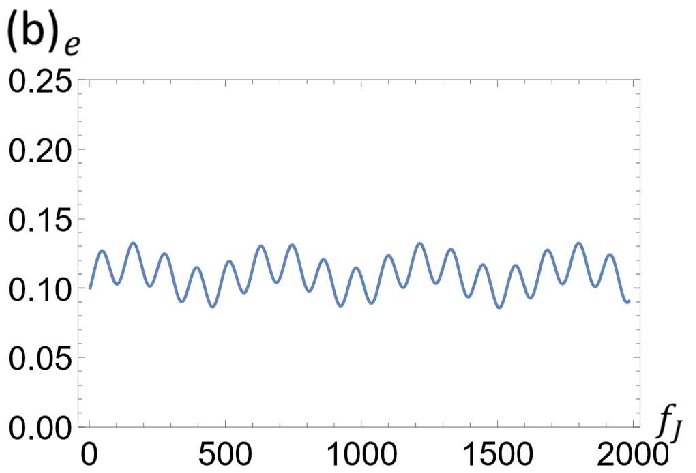}
        \end{center}
      \end{minipage}
    \end{tabular}
    \caption{Numerical results of a time evolution of the eccentricity $e$ of the asteroid with the time $f_J$ with the initial conditions $(a, e, l, g, f_J)=(a_{32}, 0.10, 2\pi/3, \pi/2, \pi)$. (a) The eccentricity is calculated by the original Hamiltonian.  (b) The eccentricity is calculated by our approximated Hamiltonian}
          \label{fig_e-timedependency-Hfj}
  \end{center}
\end{figure*}

 However, we have used a drastic approximation with a time averaging method when we evaluate the stability of the motion of asteroids by neglecting highly oscillating terms in time, even though these terms have a comparative order of the magnitude with the slowly oscillating terms. Hence, we can expect some discrepancy between the solution of the equation of motion obtained from our approximated Hamiltonian~(\ref{eq_ell_mnresonance-approximation-hamiltonian}) or (\ref{eq_ell_hamiltonian_res_yphi}) and the one obtained from the original Hamiltonian~(\ref{eq_ell-three-bodies-hamiltonian_t}) or (\ref{eq_ell-three-bodies-hamiltonian_fj}). In order to see these expected deviation we have performed numerical calculations of the equations of motion obtained from the original Hamiltonian and the approximated Hamiltonian. 

We here show 
the numerical calculations for
the shape of the Hilda triangle and the time evolution of the semimajor axis $a$ and the eccentricity $e$ of an asteroid.

The trajectories of an asteroid in the rotating coordinate are shown in Fig.~\ref{fig_triangle-Hfj}(a) and (b).
In the former
the trajectory is obtained by integrating the equation of motion with the original Hamiltonian,
while in the latter it is obtained with the approximated Hamiltonian.
We can see the shape of the Hilda triangle for both cases.
However, it seems that there is a quantitative difference between them.
In fact, the triangle is clearer in Fig.~\ref{fig_triangle-Hfj}(a).
This difference may be caused by the difference in the eccentricity $e$ discussed below.

We also present the time evolution of the semimajor axis $a$ obtained with the original Hamiltonian in Fig.~\ref{fig_a-timedependency-Hfj}(a) and that obtained with the approximated Hamiltonian in Fig.~\ref{fig_a-timedependency-Hfj}(b).
The red line in Fig.~\ref{fig_a-timedependency-Hfj}(a) is a time-averaged semimajor axis over the period $4\pi$, which agrees well with the result shown in Fig.~\ref{fig_a-timedependency-Hfj}(b).
In spite of our drastic approximation, it seems to us that our approximated results show a reasonable agreement with the one obtained by the original Hamiltonian.

On the other hand, we have seen somewhat larger discrepancy in the results for the eccentricity $e$, which are shown in Fig.~\ref{fig_e-timedependency-Hfj}(a) for the original Hamiltonian, and Fig.~\ref{fig_e-timedependency-Hfj}(b) for the approximated Hamiltonian.  A possible explanation of this discrepancy may be that we have neglected the contributions with higher power in the series expansion of the semimajor axis of the asteroid of which value is not small enough to compare with 1. We need to continue to clarify the origin of the discrepancy.

Nevertheless, it seems to us that our approximation scheme leads us to the reasonable argument of the origin of the stability of the asteroids around the 3:2 resonance point, because of the fact that we can observe the stable Hilda triangle even though these asteroids are evolving around the resonance point. Indeed, the single-resonance Hamiltonian guaranties the integrability of the system.  In other word, no mater how improve the approximation scheme better than our approximation scheme, the existence of the Hilda triangle strongly suggests that the original Hamiltonian is well-approximated by a single-resonance Hamiltonian.

\section{Summary}
\label{concluding}

 Let us summarize the results shown in this paper.  We have analyzed the motion of an asteroid of the solar system as an example of the Sun-Jupiter-asteroid elliptic restricted three-body problem around the 3:2, the 3:1, and the 2:1 resonances with respect to the orbital period of Jupiter. At the 3:2 resonance, we see a distribution of many asteroids that are known as Hilda asteroids. On the other hand, at the 3:1 and the 2:1 resonances there is no distribution of asteroids, which are known as Kirkwood gaps. Our main interest was to explain the reason of this difference in the distribution in spite of the fact that these asteroids are both influenced by the resonance with the motion of Jupiter. To find this reason we focus on the influence of the eccentricity of Jupiter on the motion of the asteroid.  

We have analyzed this by using following approximation schemes: 

(A) the perturbation analysis by neglecting small contributions in series expansion of the Hamiltonian in terms of several small parameters, 

(B) the time-average approximation by neglecting highly oscillating contributions in the Hamiltonian. 

The approximation scheme (B) was necessary for our treatment, because thanks to this procedure we could reduce the time-dependent Hamiltonian with two degrees of freedom to a time-independent Hamiltonian with two degrees of freedom.  Because of this reduction, we could analyze the stability of the motion of the asteroid in terms of the Poincar\'e surface of section.

We have treated the true anomaly of Jupiter $f_J$ as ``time" instead of the actual time $t$. This treatment is especially powerful for the elliptic restricted three-body problem, since this treatment simplify the equations of motion.

The most important result obtained in this paper is that we have found that there is a significant difference between the 3:2 resonance and the 3:1 and the 2:1 resonances. Indeed, the approximated Hamiltonian for the 3:2 resonance reduces to a single-resonance Hamiltonian which is integrable, while the approximated Hamiltonian for the 3:1 and the 2:1 resonances both reduces to a double-resonance Hamiltonian which is nonintegrable.
Hence, due to this result, we could expect regular motion of the asteroid for the 3:2 resonance, and chaotic motion for the 3:1 and the 2:1 resonances. This prediction has been verified by constructing the Poincar\'e surfaces of section using numerical integration of the equations of motion obtained by our approximated Hamiltonian. 

This result is consistent to the fact that we see a distribution of many asteroids that form the Hilda triangle in the actual observation around the 3:2 resonance, while there is no distribution of asteroids around the 3:1 and the 2:1 resonances in Kirkwood gaps.

Furthermore, we have discussed the validity of our approximation by comparison of the results obtained by our approximated Hamiltonian to the results  obtained by the original Hamiltonian before performing  the approximations. In our time-average approximation (B) mentioned above we have neglected rapid oscillating contributions, even though these terms have comparable value as compared with the slow oscillating contributions that are retained in after our approximated Hamiltonian. Because of such  a drastic approximation we could expect some discrepancy in the results obtained by the original Hamiltonian and by our approximated Hamiltonian.  Nevertheless, our approximated Hamiltonian has led to the Hilda triangle that has been reasonably agree with the Hilda triangle obtained by the original Hamiltonian for the 3:2 resonance. This has also been the case for the time evolution of the semimajor axis of the asteroid. 

 However, our approximation has not led a good agreement for the time evolution of the  eccentricity of the asteroid.  As mentioned in the previous section, a possible explanation of this discrepancy may be that we have neglected the contributions with higher power in the series expansion of the semimajor axis of the asteroid of which value is not small enough to compare with 1. Hence, we need more precise calculations to clarify the discrepancy. 
 However, because of the fact that we can observe stable asteroids around the 3:2 resonance as the Hilda triangle, while we cannot observe observe asteroids around the 3:1 and the 2:1 resonances in Kirkwood gaps, it seems to us that our main conclusion on the reduction of the Hamiltonian to the single-resonance Hamiltonian around the 3:2 resonance, while the reduction of the Hamiltonian to the multiple-resonance Hamiltonian around the 3:1 or the 2:1 resonances remains  a correct assertion to understand the stability of the asteroids in the the Hilda triangle in spite of the fact that the 3:2 resonance with Jupiter is a significant effect to the evolution of the asteroids.

\bmhead{Acknowledgments}

We thank S. Tanaka, K. Kanki and S. Garmon for many discussions and suggestions. We also thank T. Mizuguchi and T. Konishi for fruitful suggestions. This work was supported by JST, the establishment of university fellowships towards the creation of science technology innovation, Grant Number JPMJFS 2138.

\begin{appendices}

\section{The derivation of Eq.~(\ref{myuvemn})}
\label{section_d}

To expand $ \cos i_r \theta \cdot \cos j_r f \cdot \cos k_r f_J $ for $k_r = 0,1,2,...$, we first obtain
\begin{eqnarray}
\cos j_r f&=&\cos j_r l+j_r e\{\cos{(j_r +1)l}-\cos{(j_r -1)l}\} \nonumber \\
& &+e^2\{A_{f,j_r}\cos{(j_r +2)l}+B_{f,j_r}\cos{j_r l}+C_{f,j_r}\cos{(j_r -2)l}\} \nonumber \\
& &\hspace{7pt}+O(e^3), \label{eq_cosjrf-el} \\
\sin j_r f&=&\sin j_r l+j_r e\{\sin{(j_r +1)l}-\sin{(j_r -1)l}\} \nonumber \\
& &+e^2\{A_{f,j_r}\sin{(j_r +2)l}+B_{f,j_r}\sin{j_r l}+C_{f,j_r}\sin{(j_r -2)l}\} \nonumber \\
& &\hspace{7pt}+O(e^3), \label{eq_sinjrf-el} \\
\cos i_r \theta&=&\cos i_r (g+l)+i_r e\{\cos [i_r g+(i_r +1)l] -\cos [i_r g+(i_r -1)l]\} \nonumber \\
& &+e^2\{A_{f,i_r}\cos [i_r g+(i_r +2)l] +B_{f,i_r}\cos{i_r (g+l)} \nonumber \\
& &\hspace{7pt}+C_{f,i_r}\cos [i_r g+(i_r -2)l]\}+O(e^3), \label{eq_cosirtheta-elg} \\
A_{f,i_r}&=&\frac{1}{2}i_r^2+\frac{5}{8}i_r, \label{eq_afir} \\
B_{f,i_r}&=&-i_r^2, \label{eq_bfir} \\
C_{f,i_r}&=&\frac{1}{2}i_r^2-\frac{5}{8}i_r \label{eq_cfir}
\end{eqnarray}
for $ i_r, j_r = 0,1,2, ...$. These equations can be proved by the mathematical induction with Eqs.~(\ref{eq_theta-g+f})-(\ref{eq_sinf-el}). By Eqs.~(\ref{eq_cosjrf-el}) and (\ref{eq_cosirtheta-elg}), we expand $ \cos i_r \theta \cdot \cos j_r f \cdot \cos k_r f_J $ for $k_r = 0,1,2,...$ as
\begin{eqnarray}
\cos i_r \theta &\cdot& \cos j_r f \nonumber \\
&=& \frac{1}{2} \big\{ \cos [i_r g + (i_r+j_r) l] + \cos [i_r g + (i_r-j_r) l] \big\} \nonumber \\
& & +\frac{1}{2}(i_r+j_r) e \big\{ \cos [i_r g + (i_r+j_r+1) l] - \cos [i_r g + (i_r+j_r-1) l] \big\} \nonumber \\ 
& & +\frac{1}{2}(i_r-j_r) e \big\{ \cos [i_r g + (i_r-j_r+1) l] - \cos [i_r g + (i_r-j_r-1) l] \big\} \nonumber \\ 
& &+O(e^2) , \label{cosircosjr}
\end{eqnarray}\vspace{-20pt}
\begin{eqnarray}
\cos (i_r g&+&o_r l)\cdot \cos k_r f_J  \nonumber \\
&=& \frac{1}{2}\big\{ \cos (i_r g + o_r l + k_r f_J) + \cos (i_r g + o_r l -k_r f_J)\big\}
\label{cosirjrcoskr}
\end{eqnarray}
where $o_r=...,-2,-1,0,1,2,...$. Using these equations, we can represent the angle parts of the perturbation term, $ \mu V_ {\rm e}'$ in Eqs.~(\ref{eq_muvep_expansion_ftheta}) and (\ref{eq_muvep_expansion_ftheta_app}), in the form of $\cos (i_r g + o_r l \pm k_r f_J)$.


For example, we consider the terms including $ \cos (mg + nl) $. In the case of $ (i_r, j_r, k_r) = (m, m-n, 0) $, we obtain
\begin{eqnarray}
\cos m \theta &\hspace{-5pt}\cdot \hspace{-5pt}& \cos (m-n) f \cdot \cos 0 f_J \nonumber \\
&=& \frac{1}{2} \big\{ \cos [m g + (2m-n) l] + \cos (m g + n l) \big\} \nonumber \\
& & +\frac{1}{2}(2m-n) e \big\{ \cos [m g + (2m-n+1) l] - \cos [m g + (2m-n-1) l] \big\} \nonumber \\ 
& & +\frac{1}{2}n e \big\{ \cos [m g + (n+1) l] - \cos [m g + (n-1) l] \big\} \nonumber \\
& &+O(e^2).
\label{cosir=mcosjr=m-n}
\end{eqnarray}
The first term in Eq.~(\ref{cosir=mcosjr=m-n}) includes $ \cos (mg + nl) $. Similarly, in the case of $ (i_r, j_r, k_r) = (m, m-n-1,0) $, we obtain
\begin{eqnarray}
\cos m \theta &\hspace{-5pt}\cdot \hspace{-5pt}& \cos (m-n-1) f \cdot \cos 0 f_J \nonumber \\
&=& \frac{1}{2} \big\{ \cos [m g + (2m-n-1) l] + \cos [m g + (n+1) l] \big\} \nonumber \\
& & +\frac{1}{2}(2m-n-1) e \big\{ \cos [m g + (2m-n) l] - \cos [m g + (2m-n-2) l] \big\} \nonumber \\
& & +\frac{1}{2}(n+1) e \big\{ \cos [m g + (n+2) l] - \cos (m g + n l) \big\}+O(e^2).
\label{cosir=mcosjr=m-n-1}
\end{eqnarray}
The third term in Eq.~(\ref{cosir=mcosjr=m-n-1}) includes $ \cos (mg + nl) $. In the same way as Eqs.~(\ref{cosir=mcosjr=m-n}) and (\ref{cosir=mcosjr=m-n-1}), $ \cos i_r \theta \cdot \cos j_r f \cdot \cos k_r f_J $ for $ (i_r, j_r, k_r) = (1,1,1), (1,0,1) $ includes $ \cos (g + f_J) $ and $ \cos i_r \theta \cdot \cos j_r f \cdot \cos k_r f_J $ for $(i_r, j_r, k_r) = (m-1 , m-n-1,1)$ includes $ \cos [mg + nl- (g + f_J) ] $.

In the first-order resonance ($m-n=1$), we consider the long-period terms including $ \cos (mg + nl)$, $ \cos (g + f_J)$ and $ \cos [mg + nl-(g + f_J) ]$ and ignore long-period terms because of the approximation ignoring higher orders in $\mu, e_J, e$ and $a$ in Section 3. From Eq.~(\ref{eq_muvep_expansion_ftheta}), the orders in $e_J, a$ and $e$ in the term including $ \cos (m_r g + n_r l + \alpha_r f_J) $ ($ m_r, n_r$ and $\alpha_r $ are integers) increase as each of the absolute values of $ m_r $, $ n_r $ and $\alpha_r $ becomes larger. For example, the coefficient of $\cos 3(mg+nl)$ in Eq.~(\ref{eq_muvep_expansion_ftheta}) is $O(\mu a^{3m} e^{3(m-n)})$. This is much smaller than $\mu a^m e^{m-n}$ which is the order of the coefficient of $\cos (mg+nl)$ in Eq.~(\ref{cpmn}).

\section{Coefficient of $\cos (i\theta)$ in $P_i (\cos \theta)$}
\label{section_e}

Since Legendre polynomial $P_i(x)$ is
\begin{eqnarray}
P_i(x)&=&\frac{1}{2^i\cdot i!}\cdot \frac{d^i}{dx^i}(x^2-1)^i, \label{eq_app_Legendre-Pix} \\
(x^2-1)^i&=&\sum_{j=0}^i \hspace{0pt} _i{\rm C}_{j} (-1)^j x^{2(i-j)}, \label{eq_app_(x^2-1)^i}
\end{eqnarray}
we can obtain
\begin{eqnarray}
P_i(\cos \theta) =\frac{(2i)!}{2^i\cdot (i!)^2}\cos^i \theta +\frac{1}{2^i\cdot i!}\sum_{j=1}^i \hspace{0pt} _i{\rm C}_{j}(-1)^j\cdot \frac{d^i}{dx^i}x^{2(i-j)}\mid_{x=\cos \theta} , \label{eq_app_Legendre-Picostheta}
\end{eqnarray}
where $P_{0}(\cos \theta)=1$, $P_{1}(\cos \theta)= \cos \theta$, $P_{2}(\cos \theta)= \frac{1}{2}(3 \cos^2 \theta -1)=\frac{1}{4}(3 \cos 2\theta +1)$ and $P_{3}(\cos \theta)= \frac{1}{8}(5 \cos 3\theta +3\cos \theta)$. Here the coefficient of $\cos (i\theta)$ in $\cos^i \theta$ is $1/2^{i-1}$ since $\cos^i \theta$ is expanded by
\begin{eqnarray}
\cos^i \theta=
  \begin{cases}
   \frac{1}{2^{i-1}}\sum_{m'=0}^{(i-1)/2}\hspace{0pt} _i{\rm C}_{m'}\cos {(i-2m')\theta} & (i {\rm : odd \hspace{5pt} number)} \\
   \frac{1}{2^{i-1}}\sum_{m'=0}^{i/2-1}\hspace{0pt} _i{\rm C}_{m'}\cos {(i-2m')\theta}+\frac{1}{2^i} _i{\rm C}_{\frac{i}{2}} & (i {\rm : even \hspace{5pt} number)}.
  \end{cases}\hspace{20pt} \label{eq_app_costheta^i}
\end{eqnarray}
From the first term of Eq.~(\ref{eq_app_Legendre-Picostheta}) and Eq.~(\ref{eq_app_costheta^i}), we obtain the coefficient of $\cos (i\theta)$ in $P_i (\cos \theta)$,
\begin{eqnarray}
p_{\theta,i}=
  \begin{cases}
   \frac{(2i)!}{2^{2i-1}(i!)^2} & (i=1,2,...) \\
   1 & (i=0).
  \end{cases} \label{eq_app_pthetai}
\end{eqnarray}



\section{Canonical transformation in Section \ref{reduction}}
\label{section_f}

The canonical transformation from the Delaunay variables to the variables $(y_1,y_2,\phi_1, \phi_2)$ in Eqs.~(\ref{eq_y1})-(\ref{eq_phi2}) is obtained by the generating function,
\begin{equation}
W_3(l,g,y_1,y_2,f_J)=
  \begin{cases}
   (y_1+\tilde{y}_{mn}) (mg+nl) +y_2 (g+f_J-\pi) & (c_{mn}>0) \\
   (y_1+\tilde{y}_{mn}) (mg+nl-\pi) +y_2 (g+f_J-\pi) & (c_{mn}<0).
  \end{cases} \label{eq_gl-yphi_W}
\end{equation}
Hence, we can obtain the Hamiltonian (\ref{eq_newH}),
\begin{eqnarray}
H(y_1,y_2,\phi_1, \phi_2)
  &=&H_{f_J}'(L, G, l, g, f_J)+\frac{\partial W_3}{\partial f_J} \nonumber \\
  &=&H_{f_J}'(L, G, l, g, f_J)+y_2.
\end{eqnarray}

\end{appendices}


\bibliography{2022_CMDA_asano-bibliography} 


\begin{thebibliography}{27}
\ifx \bisbn   \undefined \def \bisbn  #1{ISBN #1}\fi
\ifx \binits  \undefined \def \binits#1{#1}\fi
\ifx \bauthor  \undefined \def \bauthor#1{#1}\fi
\ifx \batitle  \undefined \def \batitle#1{#1}\fi
\ifx \bjtitle  \undefined \def \bjtitle#1{#1}\fi
\ifx \bvolume  \undefined \def \bvolume#1{\textbf{#1}}\fi
\ifx \byear  \undefined \def \byear#1{#1}\fi
\ifx \bissue  \undefined \def \bissue#1{#1}\fi
\ifx \bfpage  \undefined \def \bfpage#1{#1}\fi
\ifx \blpage  \undefined \def \blpage #1{#1}\fi
\ifx \burl  \undefined \def \burl#1{\textsf{#1}}\fi
\ifx \doiurl  \undefined \def \doiurl#1{\url{https://doi.org/#1}}\fi
\ifx \betal  \undefined \def \betal{\textit{et al.}}\fi
\ifx \binstitute  \undefined \def \binstitute#1{#1}\fi
\ifx \binstitutionaled  \undefined \def \binstitutionaled#1{#1}\fi
\ifx \bctitle  \undefined \def \bctitle#1{#1}\fi
\ifx \beditor  \undefined \def \beditor#1{#1}\fi
\ifx \bpublisher  \undefined \def \bpublisher#1{#1}\fi
\ifx \bbtitle  \undefined \def \bbtitle#1{#1}\fi
\ifx \bedition  \undefined \def \bedition#1{#1}\fi
\ifx \bseriesno  \undefined \def \bseriesno#1{#1}\fi
\ifx \blocation  \undefined \def \blocation#1{#1}\fi
\ifx \bsertitle  \undefined \def \bsertitle#1{#1}\fi
\ifx \bsnm \undefined \def \bsnm#1{#1}\fi
\ifx \bsuffix \undefined \def \bsuffix#1{#1}\fi
\ifx \bparticle \undefined \def \bparticle#1{#1}\fi
\ifx \barticle \undefined \def \barticle#1{#1}\fi
\bibcommenthead
\ifx \bconfdate \undefined \def \bconfdate #1{#1}\fi
\ifx \botherref \undefined \def \botherref #1{#1}\fi
\ifx \url \undefined \def \url#1{\textsf{#1}}\fi
\ifx \bchapter \undefined \def \bchapter#1{#1}\fi
\ifx \bbook \undefined \def \bbook#1{#1}\fi
\ifx \bcomment \undefined \def \bcomment#1{#1}\fi
\ifx \oauthor \undefined \def \oauthor#1{#1}\fi
\ifx \citeauthoryear \undefined \def \citeauthoryear#1{#1}\fi
\ifx \endbibitem  \undefined \def \endbibitem {}\fi
\ifx \bconflocation  \undefined \def \bconflocation#1{#1}\fi
\ifx \arxivurl  \undefined \def \arxivurl#1{\textsf{#1}}\fi
\csname PreBibitemsHook\endcsname

\bibitem{a}
\begin{barticle}
\bauthor{\bsnm{Petrosky}, \binits{T.}},
\bauthor{\bsnm{Noba}, \binits{K.}}:
\batitle{Theoretical {A}nalysis of {A}steroid {B}elt {B}ased on the
  {L}iouvillian {D}ynamics}.
\bjtitle{Butsuri}
\bvolume{72},
\bfpage{121}--\blpage{126}
(\byear{2017})
\end{barticle}
\endbibitem

\bibitem{m}
\begin{bbook}
\bauthor{\bsnm{Kirkwood}, \binits{D.}}:
\bbtitle{Meteoric Astronomy: A Treatise on Shooting-Stars, Fireballs, and
  Aerolites},
pp. \bfpage{105}--\blpage{111}.
\bpublisher{J. B. Lippincott},
\blocation{Philadelphia}
(\byear{1867})
\end{bbook}
\endbibitem

\bibitem{ab}
\begin{barticle}
\bauthor{\bsnm{Moons}, \binits{M.}}:
\batitle{Review of the dynamics in the {K}irkwood gaps}.
\bjtitle{Celest. Mech. Dyn. Astron.}
\bvolume{65},
\bfpage{175}--\blpage{204}
(\byear{1996})
\end{barticle}
\endbibitem

\bibitem{n}
\begin{barticle}
\bauthor{\bsnm{Wisdom}, \binits{J.}}:
\batitle{A perturbative treatment of motion near the 3/1 commensurability}.
\bjtitle{Icarus}
\bvolume{63},
\bfpage{272}--\blpage{289}
(\byear{1985})
\end{barticle}
\endbibitem

\bibitem{ac}
\begin{barticle}
\bauthor{\bsnm{Wisdom}, \binits{J.}}:
\batitle{Chaotic behavior and the origin of the 31 {K}irkwood gap}.
\bjtitle{Icarus}
\bvolume{56},
\bfpage{51}--\blpage{74}
(\byear{1983})
\end{barticle}
\endbibitem

\bibitem{s}
\begin{barticle}
\bauthor{\bsnm{Wisdom}, \binits{J.}}:
\batitle{The origin of the {K}irkwood gaps: a mapping for asteroidal motion
  near the 3/1 commensurability}.
\bjtitle{Astron. J.}
\bvolume{87},
\bfpage{577}--\blpage{593}
(\byear{1982})
\end{barticle}
\endbibitem

\bibitem{j}
\begin{barticle}
\bauthor{\bsnm{Wisdom}, \binits{J.}}:
\batitle{Urey price lecture: Chaotic dynamics in the solar system}.
\bjtitle{Icarus}
\bvolume{72},
\bfpage{241}--\blpage{275}
(\byear{1987})
\end{barticle}
\endbibitem

\bibitem{ag}
\begin{barticle}
\bauthor{\bsnm{Yoshikawa}, \binits{M.}}:
\batitle{A survey on the motion of asteroids in commensurabilities with
  {J}upiter}.
\bjtitle{Astron. Astrophys.}
\bvolume{213},
\bfpage{436}--\blpage{458}
(\byear{1989})
\end{barticle}
\endbibitem

\bibitem{r}
\begin{barticle}
\bauthor{\bsnm{Murray}, \binits{C.D.}}:
\batitle{Structure of the 2/1 and 3/2 {J}ovian {R}esonances}.
\bjtitle{Icarus}
\bvolume{65},
\bfpage{70}--\blpage{82}
(\byear{1986})
\end{barticle}
\endbibitem

\bibitem{ah}
\begin{barticle}
\bauthor{\bsnm{Ferraz-Mello}, \binits{S.}}:
\batitle{The high-eccentricity libration of the {H}ildas}.
\bjtitle{Astron. J.}
\bvolume{96},
\bfpage{400}--\blpage{408}
(\byear{1988})
\end{barticle}
\endbibitem

\bibitem{u}
\begin{barticle}
\bauthor{\bsnm{Szebehely}, \binits{V.}}:
\batitle{On the problem of three bodies in a plane}.
\bjtitle{Math. Mag.}
\bvolume{26},
\bfpage{59}
(\byear{1952})
\end{barticle}
\endbibitem

\bibitem{w}
\begin{barticle}
\bauthor{\bsnm{Brouwer}, \binits{D.}}:
\batitle{The {P}roblem of the {K}irkwood {G}aps in the {A}steroid {B}elt}.
\bjtitle{Astron. J.}
\bvolume{68},
\bfpage{152}--\blpage{158}
(\byear{1963})
\end{barticle}
\endbibitem

\bibitem{aa}
\begin{barticle}
\bauthor{\bsnm{Schubart}, \binits{J.}}:
\batitle{Long-period effects in the motion of {H}ilda-type planets}.
\bjtitle{Astron. J.}
\bvolume{73},
\bfpage{99}--\blpage{103}
(\byear{1968})
\end{barticle}
\endbibitem

\bibitem{ae}
\begin{botherref}
\oauthor{\bsnm{Schubart}, \binits{J.}}:
Long-period effects in nearly commensurable cases of the restricted three-body
  problem.
SAO Spec. Rpr.
\textbf{149}
(1964)
\end{botherref}
\endbibitem

\bibitem{t}
\begin{barticle}
\bauthor{\bsnm{Ford}, \binits{J.}},
\bauthor{\bsnm{Lunsford}, \binits{G.H.}}:
\batitle{Stochastic {B}ehavior of {R}esonant {N}early {L}inear {O}scillator
  {S}ystems in the {L}imit of {Z}ero {N}onlinear {C}oupling}.
\bjtitle{Phys. Rev. A}
\bvolume{1},
\bfpage{59}--\blpage{70}
(\byear{1970})
\end{barticle}
\endbibitem

\bibitem{ai}
\begin{barticle}
\bauthor{\bsnm{Ferraz-Mello}, \binits{S.}}:
\batitle{Dynamics of the asteroidal 2/1 resonance}.
\bjtitle{Astron. J.}
\bvolume{108},
\bfpage{2330}--\blpage{2337}
(\byear{1994})
\end{barticle}
\endbibitem

\bibitem{ad}
\begin{barticle}
\bauthor{\bsnm{Giffen}, \binits{R.}}:
\batitle{A study of {C}ommensurable {M}otion in the {A}steroid {B}elt}.
\bjtitle{Astron. Astrophys.}
\bvolume{23},
\bfpage{387}--\blpage{403}
(\byear{1973})
\end{barticle}
\endbibitem

\bibitem{l}
\begin{bbook}
\bauthor{\bsnm{Reichl}, \binits{L.E.}}:
\bbtitle{The Transition to Chaos}.
\bpublisher{Springer},
\blocation{Berlin, Heidelberg}
(\byear{2004})
\end{bbook}
\endbibitem

\bibitem{af}
\begin{barticle}
\bauthor{\bsnm{Froeschl\'e}, \binits{C.}},
\bauthor{\bsnm{Scholl}, \binits{H.}}:
\batitle{On the {D}ynamical {T}opology of the {K}irkwood {G}aps}.
\bjtitle{Astron. Astrophys.}
\bvolume{48},
\bfpage{389}--\blpage{393}
(\byear{1976})
\end{barticle}
\endbibitem

\bibitem{c}
\begin{bbook}
\bauthor{\bsnm{Szebehely}, \binits{V.}}:
\bbtitle{Theory of Orbits}.
\bpublisher{Academic Press},
\blocation{New York, San Francisco, London}
(\byear{1967})
\end{bbook}
\endbibitem

\bibitem{f}
\begin{bbook}
\bauthor{\bsnm{Murray}, \binits{C.D.}},
\bauthor{\bsnm{Dermott}, \binits{S.F.}}:
\bbtitle{Solar System Dynamics}.
\bpublisher{Cambridge university press},
\blocation{Cambridge}
(\byear{1999})
\end{bbook}
\endbibitem

\bibitem{v}
\begin{barticle}
\bauthor{\bsnm{Szebehely}, \binits{V.}},
\bauthor{\bsnm{Giacaglia}, \binits{G.}}:
\batitle{On the {E}lliptic {R}estricted {P}roblem of {T}hree {B}odies}.
\bjtitle{Astron. J.}
\bvolume{69},
\bfpage{230}--\blpage{235}
(\byear{1964})
\end{barticle}
\endbibitem

\bibitem{d}
\begin{barticle}
\bauthor{\bsnm{Petrosky}, \binits{T.}}:
\batitle{{L}evel {R}epulsion and {T}hreefold {D}egeneracy of {E}igenstates of
  the {L}iouvillian in the {K}irkwood {G}aps of {A}steroid {B}elt}.
\bjtitle{Prog. Theor. Phys.}
\bvolume{125},
\bfpage{411}--\blpage{434}
(\byear{2011})
\end{barticle}
\endbibitem

\bibitem{g}
\begin{bbook}
\bauthor{\bsnm{Greiner}, \binits{W.}}:
\bbtitle{Quantum Mechanics; An Introduction}.
\bpublisher{Springer},
\blocation{Berlin, Heidelberg}
(\byear{2009})
\end{bbook}
\endbibitem

\bibitem{x}
\begin{barticle}
\bauthor{\bsnm{Chirikov}, \binits{B.}}:
\batitle{A universal instability of many-dimensional oscillator systems}.
\bjtitle{Phys. Rep.}
\bvolume{52},
\bfpage{263}--\blpage{379}
(\byear{1979})
\end{barticle}
\endbibitem

\bibitem{o}
\begin{botherref}
\oauthor{\bsnm{Arnold}, \binits{V.I.}},
\oauthor{\bsnm{Avez}, \binits{A.}}:
Ergodic Problems of Classical Mechanics,
New York, Benjamin
(1968)
\end{botherref}
\endbibitem

\bibitem{p}
\begin{bbook}
\bauthor{\bsnm{Arnold}, \binits{V.I.}}:
\bbtitle{Mathematical Methods of Classical Mechanics}.
\bpublisher{Springer},
\blocation{Berlin, Heidelberg}
(\byear{1989})
\end{bbook}
\endbibitem

\end{thebibliography}


\end{document}